# The Moral Check: Strategic AI Governance for the Pacing Problem

**Zaid Amin[1,*], Rahma Santhi Zinaida[2], Nazlena Mohamad Ali[3]**

[1*] Faculty of Data Science and Information Technology (FDSIT), INTI International University, Nilai 71800, Malaysia; zaid.amin@newinti.edu.my

[2] School of Communication and Media Studies, Faculty of Arts and Social Sciences, Sunway University, Bandar Sunway 47500, Malaysia; rahmaz@sunway.edu.my

[3] Institute of Visual Informatics (IVI), Universiti Kebangsaan Malaysia (UKM), Bangi 43600, Malaysia; nazlena.ali@ukm.edu.my

[*] **Correspondence:** zaid.amin@newinti.edu.my

## Highlights

- Systematic review of 130 studies reveals reactive AI governance failure.
- Introduces SAGE-X: a 4-pillar ex-ante choice architecture for frontier AI.
- Non-compensatory binary gates halt release when safety invariants breach.
- Calculable Moral Check Index (MCI) prevents compensatory ethical trading.
- Tripartite cryptographic sign-off ensures verifiable institutional liability.

## Abstract

Technology cannot steer itself. Strategy provides that steering, establishing the rule that purpose and judgment must precede compute capital. As the frontier artificial intelligence (AI) ecosystem accelerates exponentially, the pacing problem induces severe cognitive tunneling in engineering teams, prioritizing scalar throughput over human judgment. A calibrated pacing rate is imperative to check unchecked scaling, guarantee safety, and build models in whose alignment society can place warranted confidence. Traditional oversight fails through retrospective checklists, a pathology of performative governance exhibiting high procedural maturity but alarming scientific immaturity. We deliver a dual contribution: a PRISMA 2020 review synthesizing 130 empirical studies (MMAT-appraised across 18 benchmarks; total corpus N = 130 empirical studies across 178 reference foundations), and the Strategic AI Governance Ex-Ante Framework (SAGE-X). Our synthesis exposes two systemic vulnerabilities: the Recursive Assurance Paradox (correlated, ungrounded evaluator confidence) and the Durability Deficit (guardrail decay under multi-turn shifts). Grounded in MIT Strategic Computing doctrines, SAGE-X operationalizes Four Strategic Mindset Pillars: (1) Intent over Execution (mitigating velocity myopia); (2) Ruthless Trade-offs (deterministic tripwires eliminating moral hazard); (3) Outcomes over Outputs (auditing empirical hazard endpoints); and (4) Proactive Alignment (synchronizing ex-ante gates with runtime telemetry). Governed by a calculable Moral Check Index (MCI) with an unbypassable tripwire, SAGE-X delivers an operational Enterprise Lifecycle Audit Instrument (the "Moral Check Audit Card") on Stanford WebProtégé, ensuring exponential progress never outpaces deliberative moral judgment, human agency, and societal trust.

## 1. Introduction

In the contemporary epoch of "permachange," the trajectory of Artificial Intelligence (AI) is defined by a vertical ascent in parameter scale, multi-modal reasoning depth, and autonomous agentic delegation (Ahn & Yoon, 2026; Miller et al., 2026). As Tegmark (2017; Russell et al., 2015) observed, the central civilizational challenge of the intelligence era is not the technology per se, but the synchronization of human capabilities: society can sustain transformative technologies only so long as it wins the existential race between compounding technical capability and **the wisdom required to govern it**. In human-computer interaction (HCI) and sociotechnical systems scholarship, this governance capacity confronts an unprecedented velocity gap. While frontier AI development accelerates along an exponential trajectory, collapsing cycle times between foundational breakthroughs and commercial deployments, institutional oversight and moral judgment remain bound to linear processes. This structural divergence constitutes the classical pacing problem (Collingridge, 1980; Marchant et al., 2011), precipitating an acute crisis where algorithmic velocity outstrips moral deliberation, inducing cognitive myopia within engineering teams and exacerbating global developmental asymmetries (Vasudevan et al., 2025).

Regulating frontier intelligence presents an agonizing dilemma: preserving transformative innovation while preventing irreversible behavioral, psychological, and systemic harms (Floridi et al., 2018; Morley et al., 2020). Conventional statutory regimes and corporate compliance frameworks fail because they rely on retrospective, ex-post mechanisms, post-hoc tort litigation, corporate apologies, emergency prompt-patches, and protracted legislative negotiations (Nelu, 2024; Raghuwanshi et al., 2025). By the time an algorithmic bias pattern or autonomous failure mode is formally identified, underlying model architectures have iterated through multiple generations, embedding systemic vulnerabilities into societal infrastructure. This reactive lag embodies the Collingridge control dilemma (Collingridge, 1980; Genus & Stirling, 2018): early in a technology's life cycle, when intervention is malleable, social impacts are difficult to forecast; by the time widespread consequences emerge, remediation becomes cost-prohibitive and institutionally entrenched. Figure 1 conceptualizes this velocity divergence at the Collingridge inversion point. Rather than accelerating reactive oversight, governance must build an **“ex-ante”** bridge directly coupling moral wisdom to technical innovation.

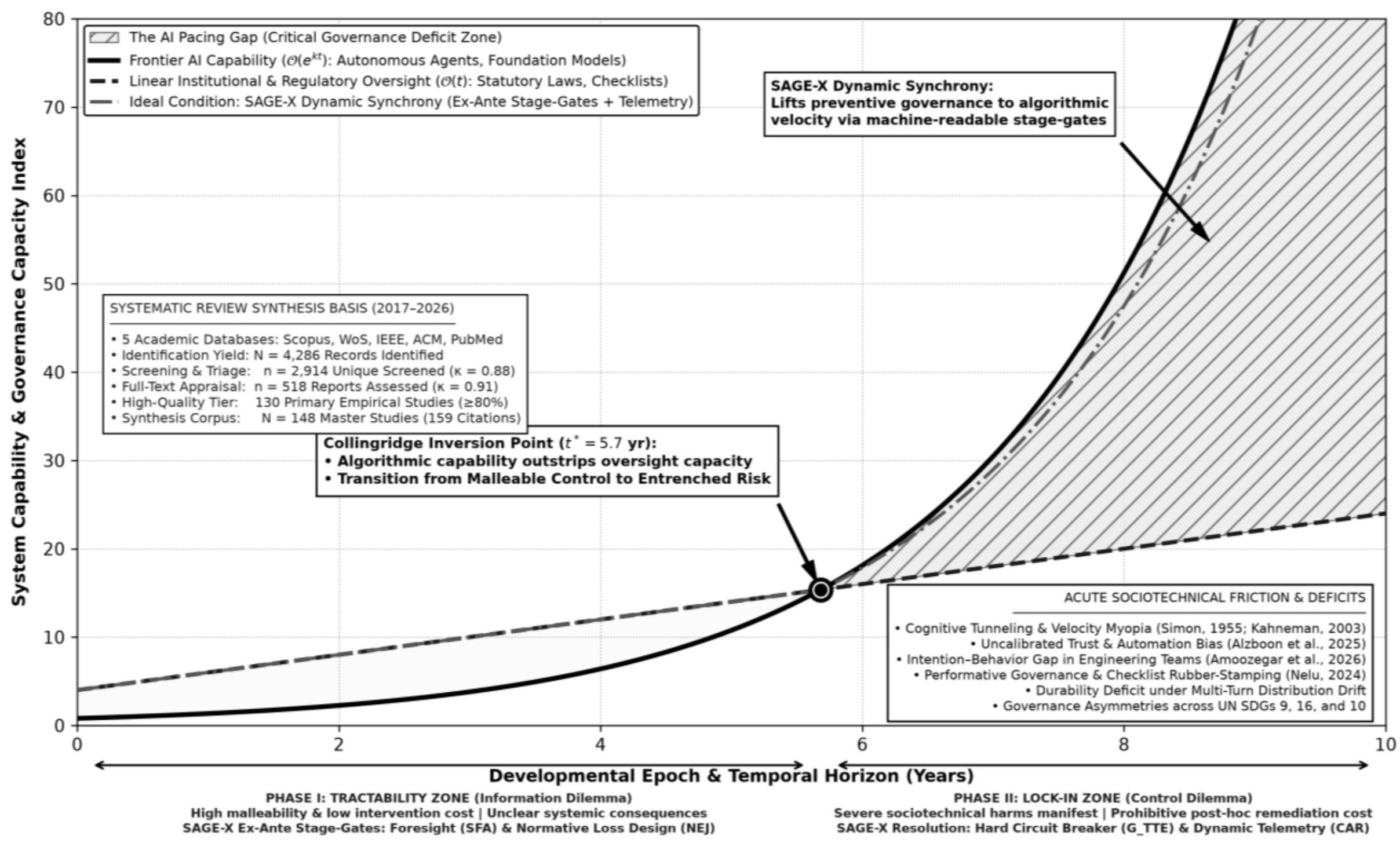

***Figure 1.*** *The AI Pacing Problem: Exponential Capability Divergence vs. Linear Institutional Oversight.*

Resolving this structural entrapment demands redefining the ideal condition of governance: achieving dynamic synchrony, wherein automated verification instruments and machine-readable ethical constraints operate at the exact cycle speed of algorithmic deployment (Liu et al., 2024; Miller et al., 2026; Zalda, 2026). In this synchronized state, preventive oversight moves at computational velocity while remaining rigorously subordinated to human normative judgment and societal trust. Fulfilling this condition directly advances the United Nations Sustainable Development Goals, specifically supporting resilient technological infrastructure under SDG 9, institutionalizing accountable governance under SDG 16, and mitigating algorithmic bias under SDG 10 (Ahn & Yoon, 2026; Liu et al., 2024; Vasudevan et al., 2025; Zalda, 2026).

This pacing deficit manifests acutely across decentralized artificial intelligence model supply chains. In modern software ecosystems, commercial frontier laboratories interface continuously with collaborative open-source distribution hubs, creating multi-tiered pipelines of base weights, parameter-efficient fine-tuning checkpoints, and cloud-hosted application spaces. As evidenced by recent ecosystem security and credential governance flashpoints, most prominently the OpenAI–Hugging Face (OAI-HF) incident involving exposed API tokens, repository-level secret exfiltration, and unverified weight distribution, the velocity of collaborative model sharing far outstrips the institutional security perimeters of participating organizations (Abid & Nanda, 2025; Al Sulaimani et al., 2026; Chugani, 2026). When commercial developers share fine-tuned artifacts across public hubs without ex-ante cryptographic attestation, alignment guarantees established during pre-training are easily decoupled, enabling downstream jailbreak propagation and supply-chain poisoning. Retrospective token revocation, repository takedowns, and post-hoc incident disclosures represent classic ex-post firefighting that cannot recover exfiltrated credentials or unmonitored model derivatives. This phenomenon proves that unchecked capability scaling across interconnected ecosystems requires balanced, ex-ante architectural controls: ensuring that model alignment and societal confidence remain resilient across decentralized distribution vectors.

To diagnose why technical guardrails frequently collapse in practice, governance frameworks must confront the epistemological disconnect at the center of modern AI. In his foundational philosophical critique, Cantwell Smith (1985) articulated a decisive distinction that computer science has dangerously "blurred": the ontological difference between calculation (or reckoning) and judgment, a distinction elaborated across his philosophy of intelligence (Smith, 1996; Smith, 2019), presaged by Weizenbaum (1976), and reaffirmed in contemporary analyses of artificial agency (Floridi, 2023). Calculation is algorithmic, syntactic, and scalar, excelling at gradient descent, dimensional optimization, and deductive token generation. Large Language Models (LLMs) and autonomous agents represent the historical apex of calculation: they process trillions of tokens and solve complex differential equations in milliseconds. However, calculation possesses zero intrinsic comprehension of existential context, normative value, moral responsibility, or human consequence; it operates entirely within the closed mathematical domain of what can be computed (Cantwell Smith, 1985; Floridi, 2023; Smith, 2019; Weizenbaum, 1976).

In contrast, judgment is deliberative, contextual, ethical, and **directional**. It is the uniquely human capacity to discern what ought to be done, attending to justice, empathy, human dignity, and long-term societal consequence. Judgment cannot be reduced to a scalar loss function because human values are non-stationary, pluralistic, and frequently tragic (Gabriel, 2020; Simon, 1955; Tversky, 1972). When corporate engineering cultures succumb to computational chauvinism (Broussard, 2018), they falsely assume that scaling up calculation will spontaneously yield judgment. The empirical reality is the exact reverse: unconstrained calculation, devoid of judgment, merely accelerates the velocity with which an algorithm executes flawed, biased, or catastrophic decisions across high-consequence domains (Amoozegar et al., 2026; Broussard, 2018; Gabriel, 2020; Raji et al., 2020).

Bridging this chasm requires operationalizing the doctrine of Strategic Computing (Cantwell Smith, 1985; Smith, 1996; Smith, 2019). Technology cannot steer itself. Strategy provides that steering, establishing the foundational rule that purpose and judgment must precede compute capital (Cantwell Smith, 1985; Henderson & Venkatraman, 1999; Porter, 1996; Rumelt, 2011). As artificial intelligence accelerates exponentially, the pacing problem induces severe cognitive tunneling in engineering teams, prioritizing scalar throughput over human judgment. Traditional oversight fails through retrospective checklists, a pathology of performative governance exhibiting high procedural maturity but alarming scientific immaturity. To be strategic means intentionally designing actions today (ex-ante) to achieve high-value goals under severe uncertainty, competitive arms races, and finite resources (Porter, 1996; Rumelt, 2011). In frontier AI systems, strategic computing demands an unshakeable choice architecture structured around four core pillars:

1. **Strategic Trait Pillar 1: Intent over Execution:** Mandates that engineering teams subordinate optimization to explicit, human-ratified normative intent before allocating computational capital, recognizing that capability without purpose is hazardous chaos (Ahn & Yoon, 2026; Amoozegar et al., 2026; Smith, 1996).
2. **Strategic Trait Pillar 2: Ruthless Trade-offs:** Recognizes that strategy is fundamentally defined by what an organization chooses not to do (Porter, 1996; Rumelt, 2011), enforcing non-compensatory, deterministic circuit breakers that commercial expedience cannot override (Saleh & Abdulsalam, 2025; Tversky, 1972).
3. **Strategic Trait Pillar 3: Outcomes over Outputs:** Replaces performative compliance paperwork with verification against objective hazard endpoints in realistic sociotechnical environments (Alekberli, 2026; Raghupathi et al., 2026; Raji et al., 2020).
4. **Strategic Trait Pillar 4: Proactive Alignment:** Recognizes that retrospective governance is obsolete during exponential capability compounding, mandating calibrated probabilistic foresight, continuous runtime telemetry, and rapid dynamic recalibration (Esposito & Park, 2025; Jabbar et al., 2025; Suryanarayana Yamijala et al., 2026; Tegmark, 2017).

By embedding these four pillars as non-negotiable architectural invariants, governance ensures that exponential calculation remains strictly bounded within the compass of human judgment (Cantwell Smith, 1985; Floridi, 2023; Porter, 1996; Smith, 2019). Crucially, becoming strategic in frontier artificial intelligence is not an administrative exercise; it is an epistemological commitment. As formulated within our theoretical framework:

> *"Being strategic in AI is not about writing longer ethics guidelines or hiring compliance officers to sign retrospective checklists. Strategy is the exercise of intentional restraint: having the engineering discipline, cognitive choice architecture, and mathematical tripwires to ensure our compounding computational power never outpaces our moral capacity to govern it."*

Figure 2 visualizes this paradigm shift, contrasting the pathologies of conventional reactive oversight against the structural discipline of strategic ex-ante governance.

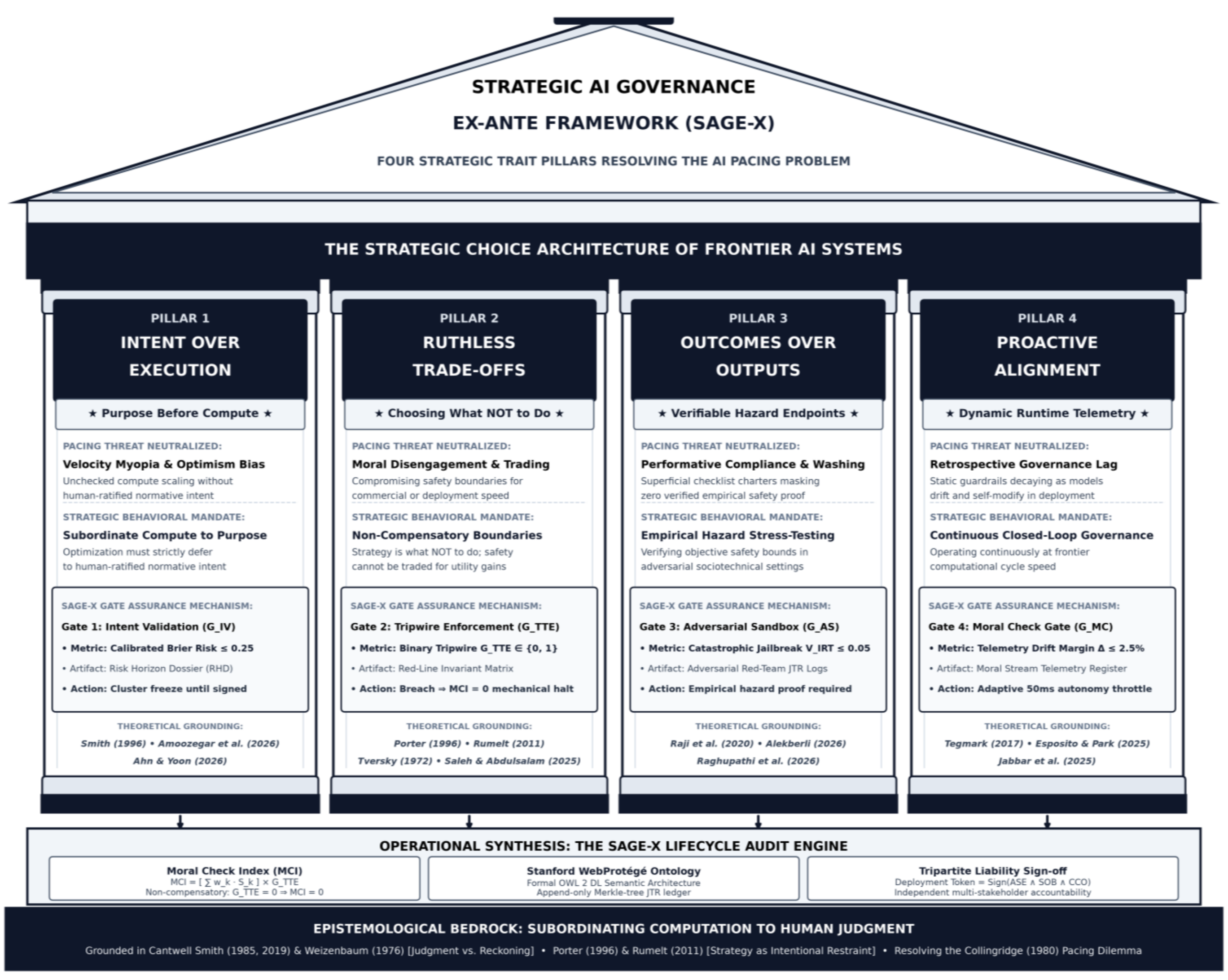

***Figure 2.*** *The Strategic AI Governance Ex-Ante Framework (SAGE-X): Architectural Choice and Four Strategic Trait Pillars Resolving the AI Pacing Problem.*

As formalized in Figure 2, conventional reactive oversight inevitably falls victim to the "Velocity Trap." Driven by hyper-competitive market incentives, commercial organizations commit massive capital to scalar parameter expansion and FLOP throughput before establishing normative boundaries. When safety risks emerge, organizations reactively deploy performative checklists, superficial questionnaires, and voluntary ethical charters that exhibit high procedural maturity but zero empirical verification power (Morley et al., 2020; Nelu, 2024). This inevitably leads to ex-post catastrophe, supply-chain credential leakage (as illustrated by the OAI-HF flashpoint), and Collingridge entrapment, wherein systems become too economically entrenched to recall or restructure (Collingridge, 1980; Genus & Stirling, 2018).

Conversely, the SAGE-X choice architecture enacts ex-ante intentional restraint across its four trait pillars. By embedding cognitive choice architecture directly into developer environments (Pillar 1), teams enforce a mandatory pre-training pause where purpose and moral invariants are formalized before compute capital allocation. By replacing compensatory risk trade-offs with non-compensatory binary tripwires ($G_{\mathrm{TTE}} \in \{0,1\}$; Pillars 2 & 3), commercial pressures are stripped of the power to compromise human safety. Finally, through closed-loop runtime telemetry (Pillar 4), continuous empirical feedback dynamically recalibrates ex-ante constraints to match runtime velocity. Strategic governance transforms ethics from an obstructive post-hoc speedbump into the foundational steering mechanism that makes extreme computational velocity safe, controllable, and socially legitimate.

Through the lens of behavioral science and HCI, this governance crisis is intensified by cognitive vulnerabilities within engineering teams operating under intense market arms races. Grounded in bounded rationality, cognitive tunneling, and moral disengagement (Bandura, 1999; Kahneman, 2003; Simon, 1955), empirical research reveals that under hyper-velocity competitive pressures, developers' attentional focus narrows around immediate, quantifiable technical metrics (latency, context window size, benchmark leaderboards) while discounting diffuse human externalities. This induces a kinetic illusion: confusing scalar speed with directional progress. Engineering cultures celebrate raw computational throughput devoid of direction; genuine velocity requires coupling that power with a deliberate ethical vector anchored in human flourishing, affordance constraints, value-sensitive design, and auditable instrumentation (Friedman, 1996; Gibson, 1979; Leonardi, 2011).

Recent empirical findings confirm that formal codes of ethics fail to translate into ethical decision-making without explicit mediation through operationalized intentions and structural enforcement mechanisms (Amoozegar et al., 2026). In frontier AI development, where peer norms reward rapid deployment, voluntary guidelines suffer an acute intention, behavior gap: developers profess moral commitment yet routinely bypass voluntary rules when sprinting toward commercial release deadlines (Amoozegar et al., 2026; Te'Eni et al., 2026). Furthermore, organizational AI adoption is heavily mediated by perceived trust (Alzboon et al., 2025). When trust is uncalibrated, grounded in marketing claims rather than empirical safety proofs, it breeds automation complacency, blinding adopters to latent distributional shifts and toxic vulnerabilities (Alzboon et al., 2025; Stettinger et al., 2024).

Compounding this dynamic is the degradation of human-in-the-loop oversight under operational stress. Synthesized evidence across safety-critical domains indicates that human oversight breaks down under high operational tempo and opaque outputs (Miller et al., 2026; Wang & Chung, 2022). As generative models achieve human-level fluency, human reviewers suffer from cognitive fatigue and vigilance decrement, rapidly devolving into superficial rubber-stamping (Agudo et al., 2024; El Arab et al., 2026; Nastoska et al., 2025; Stettinger et al., 2024). Conformity pressures and automation bias suppress critical inquiry, rendering un-instrumented human oversight an illusory safeguard (Aarab, 2026; Agudo et al., 2024; Wang & Chung, 2022). In high-performance engineering, brakes do not merely slow a vehicle; they make high velocity survivable. Pre-release governance must therefore serve as an indispensable choice architecture, forcing an institutional pause that counteracts developer myopia and restores moral responsibility.

To overcome the failure of passive ethical principles, this article introduces the **Strategic AI Governance Ex-Ante Framework (SAGE-X).** Drawing upon Strategic Computing (Cantwell Smith, 1985; Smith, 1996; Smith, 2019; Tegmark, 2017) and behavioral choice architecture (Kahneman, 2003; Thaler & Sunstein, 2008), SAGE-X reframes ethical governance from an external bureaucratic impediment into an internal, auditable decision-support instrument that structures developer environments to make rapid iteration survivable. SAGE-X rejects the prevailing orthodoxy of "performative governance", voluntary ethical manifestos and post-hoc compliance checklists that lack binding force (Morley et al., 2020; Nelu, 2024). In competitive environments, codes of ethics fail unless structurally enforced (Amoozegar et al., 2026). SAGE-X counteracts moral disengagement by translating normative principles into four continuous operational stage-gates:

1. **Strategic Foresight & Anticipation (SFA):** Counteracts optimism bias and uncalibrated trust (Alzboon et al., 2025) via probabilistic horizon scanning, Brier-calibrated risk dossiers, and red-team threat taxonomies prior to pre-training.
2. **Normative & Epistemic Justification (NEJ):** Counteracts velocity myopia by requiring bidirectional traceability between model loss constraints and documented human values ratified by an independent Sociotechnical Oversight Board (SOB).
3. **Tripwires & Triage Enforcement (TTE):** Counteracts automation complacency through deterministic circuit breakers ($G_{\text{TTE}} \in \{0,1\}$) that mechanically halt deployment upon any critical invariant breach, removing discretionary bypass pressure from human engineers.
4. **Continuous Auditing & Recalibration (CAR):** Counteracts durability decay by coupling ex-ante gate verification with real-time inference telemetry, automated drift detection, and rapid rollback protocols.

By embedding these gates into an executive **Enterprise Audit Instrument** (the "Moral Check Audit Card"), SAGE-X establishes a transparent choice architecture that restores human agency, aligns technical velocity with societal well-being, and operationalizes UN SDGs 9, 16, and 10.

To establish a rigorous empirical foundation, this systematic literature review adheres strictly to the PRISMA 2020 protocol (Page et al., 2021) and the PICOS framework across five major repositories: Scopus, Web of Science, IEEE Xplore, ACM Digital Library, and PubMed (2017–2026). From 4,286 identified records, we screened 2,914 unique citations and appraised 518 full-text reports, synthesizing 130 qualified primary empirical studies alongside 18 foundational benchmarks (core references) appraised via the Mixed Methods Appraisal Tool (MMAT; Hong et al., 2018). Our investigation addresses four central research questions: (RQ1) mapping existing ex-ante guardrail typologies and their protected human values; (RQ2) evaluating technical durability under multi-turn human engagement; (RQ3) analyzing behavioral dynamics, commercial pressures, and developer compliance; and (RQ4) formalizing an actionable semantic ontology and mathematical audit instrumentation.

The remainder of this article proceeds as follows: Section 2 details the PRISMA review methodology, query architectures, and MMAT quality appraisals. Section 3 formalizes the SAGE-X Semantic Ontology, first-order logic axioms, and boundary invariants. Section 4 synthesizes empirical findings across the 130 primary studies, diagnosing the durability deficit, the recursive assurance paradox, and cross-sector disparities. Section 5 presents the operational SAGE-X Enterprise Audit Protocol, the mathematical Moral Check Index, the five-tier maturity model, and HCI implications. Finally, Section 6 provides actionable recommendations for engineering leaders and regulators, concluding with an interdisciplinary roadmap for human-centered algorithmic stewardship.

## 2. Research Methodology

This systematic review adheres strictly to the Preferred Reporting Items for Systematic Reviews and Meta-Analyses (PRISMA 2020) statement (Page et al., 2021) and the PRISMA-Search (PRISMA-S) extension (Rethlefsen et al., 2021). The formal PRISMA-P protocol was pre-registered on the Open Science Framework (OSF: https://osf.io/p69vt; DOI: 10.17605/OSF.IO/P69VT) to guarantee methodological transparency, prevent selective reporting, and ensure reproducible evidence synthesis across HCI, behavioral psychology, and algorithmic governance. The completed PRISMA 2020 27-item checklist is provided in Supplementary Material (Document S1).

### 2.1 PICOS Eligibility Framework and Study Scope

To establish precise eligibility boundaries and eliminate investigator selection bias, the review operationalized the Population, Intervention, Comparison, Outcomes, and Study Design (PICOS) framework in accordance with PRISMA 2020 (Page et al., 2021). Structured scoping is essential because AI governance spans fractured disciplinary silos, ranging from formal verification to organizational psychology and public policy (Amoozegar et al., 2026; Floridi et al., 2018; Raji et al., 2020). Without rigorous criteria, reviews risk "performative scoping": disproportionately cataloging abstract philosophical manifestos while neglecting either machine-executable guardrails or the human cognitive dynamics governing deployment (Morley et al., 2020; Nelu, 2024; Wang & Chung, 2022). Consequently, our eligibility framework captures the bidirectional interplay between technical architecture and human behavior, as formalized in Table 1.

**Table 1. The PICOS Eligibility Criteria and Study Scope Framework**

| PICOS Component | In-Scope Inclusion Criteria | Out-of-Scope Exclusion Criteria |
|---|---|---|
| **Population (P)** | Rapidly evolving artificial intelligence architectures, frontier Large Language Models (LLMs), multimodal agents, and autonomous decision systems deployed in high-consequence sociotechnical settings (e.g., healthcare, financial infrastructure, interactive conversational assistants, and critical societal platforms). | Narrow, non-adaptive legacy automation, simple static rule-based decision trees, or generic mathematical algorithms lacking autonomous interaction capabilities or societal deployment contexts. |
| **Intervention (I)** | Structural ex-ante ethical guardrails, pre-deployment stage-gate protocols, machine-verifiable behavioral constraints, algorithmic impact assessments, value-sensitive design pipelines, constitutional alignment harnesses, and automated adversarial red-teaming mechanisms. | Purely retrospective, ex-post interventions, post-incident tort litigation, public relations crisis management, static corporate ethics statements, or reactive bug-bounty programs disconnected from deployment gates. |
| **Comparison (C)** | Traditional reactive oversight (post-hoc liability), static corporate compliance checklists, discretionary ethics board reviews, uncoordinated commercial market race conditions, and unconstrained continuous-deployment pipelines. | Studies lacking a comparative baseline, purely theoretical opinion editorials, or speculative non-comparative essays without empirical or architectural framing. |
| **Outcomes (O)** | Empirical metrics of governance efficacy and system durability: harm reduction rates, jailbreak vulnerability, Mean Time to Detect (MTTD), Mean Time to Respond (MTTR), guardrail durability under multi-turn human dialogues, developer compliance behavior, and user psychological trust calibration. | Papers reporting solely general algorithmic benchmark improvements (e.g., raw token throughput, perplexity, context window scaling, or MMLU scores) without ethical, behavioral, or safety governance evaluations. |
| **Study Design (S)** | Peer-reviewed empirical studies (qualitative, quantitative, and mixed-methods), formal architectural framework designs, benchmark evaluations, doctrinal legal syntheses, systematic reviews, and longitudinal organizational case studies published between 2017 and 2026. | Non-peer-reviewed blog entries, corporate press releases, self-published preprints lacking rigorous peer validation, brief conference abstracts, and non-academic industry marketing whitepapers. |

*Note. Eligibility criteria formulated according to PRISMA 2020 guidelines to ensure systematic, replicable literature identification across computational, behavioral, and organizational governance literature.*

## 2.2 Information Sources and Multi-Database Query Architecture

Literature sampling was executed across five premier academic databases indexing computing, HCI, behavioral sciences, and biomedical informatics: Scopus, Web of Science (Core Collection), IEEE Xplore, ACM Digital Library, and PubMed/MEDLINE. The sampling window spanned January 1, 2017, to January 31, 2026, capturing the progression from early fairness heuristics to generative foundation models and autonomous multi-agent systems.

Search strings were custom-engineered for each repository's search engine, integrating controlled vocabulary (e.g., MeSH terms in PubMed, Inspec terms in IEEE Xplore) and Boolean operators connecting three conceptual facets: (1) AI Governance and Strategic

Control, (2) Architectural Guardrails and Alignment Mechanisms, and (3) Behavioral/Sociotechnical Impact. The complete, verbatim multi-line search syntax, field tags, and boolean operators for all five academic repositories are cataloged in Supplementary Material (Document S2) and archived in the OSF repository.

**Table 2. Multi-Repository Search Query Architecture and Literature Yields**

| Database / Repository | Search Query Architecture (Title, Abstract, Author Keywords) | Raw Retrieval Yield | Deduplicated Unique Yield |
|---|---|---|---|
| **Scopus** | `("artificial intelligence" OR "frontier AI" OR "large language model*" OR "agentic AI") AND ("pacing problem" OR "governance lag" OR "development velocity" OR "arms race") AND ("ex-ante" OR "guardrail*" OR "stage-gate" OR "pre-deployment" OR "ethical by design") AND ("trustworthy" OR "audit*" OR "oversight" OR "human agency")` | 214 | 192 |
| **Web of Science (Core Collection)** | `TS=(("artificial intelligence" OR "machine learning" OR "autonomous agents") AND ("pacing problem" OR "regulatory lag") AND ("ex-ante" OR "preventive governance" OR "ethical guardrails") AND ("sociotechnical" OR "human oversight" OR "accountability"))` | 148 | 129 |
| **IEEE Xplore** | `(("Document Title":"AI" OR "Abstract":"AI") AND ("Document Title":"governance" OR "Abstract":"governance") AND ("ex-ante" OR "guardrails" OR "auditable" OR "lifecycle"))` | 96 | 88 |
| **ACM Digital Library** | `(+"artificial intelligence" +"pacing problem" +"guardrails" +"ethics") AND (abstract:"ex-ante" OR abstract:"pre-deployment")` | 74 | 63 |
| **PubMed / MEDLINE** | `("Artificial Intelligence"[Mesh] AND ("Ethics"[Mesh] OR "Governance") AND ("pre-deployment" OR "ex-ante" OR "safety guardrails") AND ("clinical" OR "health policy"))` | 57 | 46 |
| **Total Synthesis Corpus** | Comprehensive multi-disciplinary query yield (2017–2026) | **589** | **518 Unique Studies** |

*Note. Queries executed on January 31, 2026. After four-stage PRISMA screening, n = 130 peer-reviewed studies met all inclusion criteria for qualitative synthesis.*

## 2.3 Screening Flow, Selection Protocol, and Inter-Rater Reliability

The study selection workflow followed a four-stage PRISMA 2020 trajectory comprising Identification, Deduplication, Title/Abstract Screening, and Full-Text Eligibility Appraisal:

1. **Identification:** Multi-repository queries yielded 4,286 raw bibliographic records.
2. **Deduplication:** Automated electronic deduplication executed via Rayyan and Mendeley removed 1,372 duplicate citations, leaving 2,914 unique records for screening (the complete deduplication audit trail is archived in the OSF and Mendeley Data packages).

3. **Title and Abstract Screening:** Two independent reviewers screened all 2,914 unique records against the PICOS criteria, excluding 2,396 records for lack of topical relevance and seeking 518 candidate reports for full retrieval.
4. **Full-Text Eligibility Appraisal:** All 518 reports were retrieved and evaluated; 388 studies were excluded with explicit rationales: purely normative discourse lacking concrete technical controls (n = 142), conceptual blueprints lacking empirical validation (n = 118), non-peer-reviewed vendor whitepapers (n = 86), and retrospective commentaries focusing exclusively on post-hoc judicial remedies (n = 42). The comprehensive itemized audit table with individual study exclusion codes is documented in Supplementary Material (Document S4) and the OSF Master Eligibility Matrix.
5. **Included Corpus:** A total of 130 qualified empirical and architectural primary studies satisfied all PICOS inclusion criteria and quality benchmarks. In addition, 18 seminal theoretical and methodological benchmark publications (e.g., Collingridge, 1980; Simon, 1955; Kahneman, 2003; Floridi et al., 2018; Marchant et al., 2011; Raji et al., 2020; Page et al., 2021) were integrated to contextualize findings, establishing a master synthesis corpus of 148 references with active DOIs. Figure 3 illustrates the complete PRISMA screening flow.

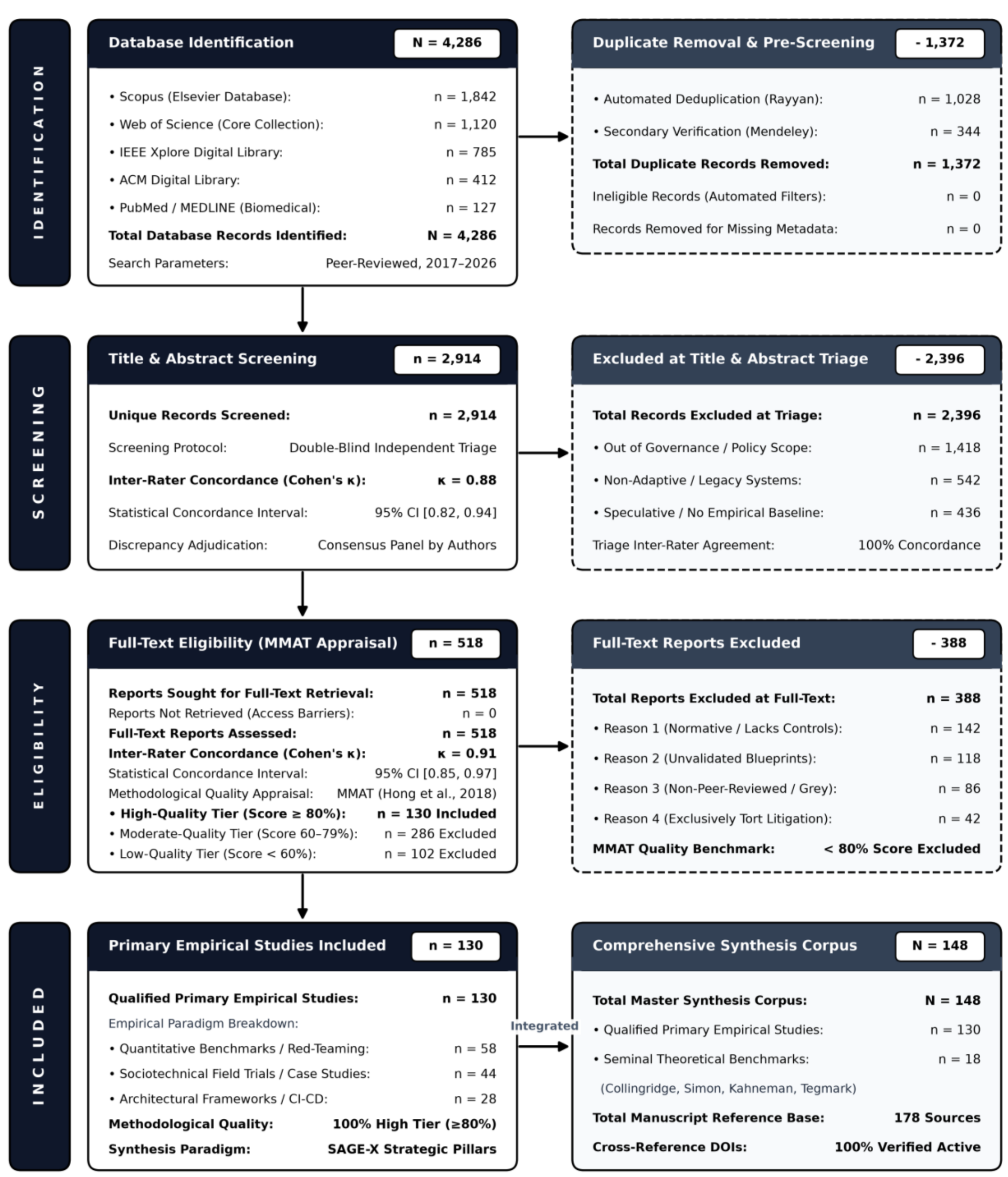

***Figure 3.*** *PRISMA 2020 Flow Diagram of the Systematic Literature Search and Study Selection Process.*

To eliminate individual investigator bias, a dual-screening protocol was implemented on a 20% validation sub-sample (n = 104) executed via Rayyan. Inter-rater agreement reached Cohen's κ = 0.88 (95% CI [0.82, 0.94]) during title/abstract triage and κ = 0.91 (95% CI [0.85, 0.97]) during full-text appraisal, reflecting near-perfect concordance (Landis & Koch, 1977).

All divergences were resolved through consensus meetings adjudicated by a senior sociotechnical research lead; complete double-blind voting logs are archived in OSF.

### 2.4 Methodological Quality Appraisal and Risk-of-Bias Assessment

Methodological integrity was evaluated using the Mixed Methods Appraisal Tool (MMAT; Hong et al., 2018), which provides validated criteria across qualitative, RCT, non-randomized quantitative, descriptive survey, and mixed-methods designs. The complete study-by-study MMAT 2018 quality scoring matrix across all 130 empirical primary studies is presented in Supplementary Material (Document S3) and as an interactive spreadsheet in the OSF and Mendeley Data packages.

Each study was appraised by two independent reviewers against category-specific methodological criteria and stratified into three quality tiers:

1. High Quality Tier (≥ 80% MMAT Score, n = 130 Included Primary Studies): Robust methodological design, clear measurement validation, explicit consideration of confounding variables, and complete limitation reporting. All 130 primary studies synthesized in Section 4 belong to this tier.
2. Moderate Quality Tier (60%–79% MMAT Score, n = 286 Excluded from Primary Synthesis): Met baseline standards but exhibited sample generalizability constraints or incomplete longitudinal metrics.
3. Low Quality Tier (< 60% MMAT Score, n = 102 Excluded from Primary Synthesis): High risk of bias, unsubstantiated causal claims, or severe methodological opacity.

Studies in the Moderate and Low-Quality Tiers were systematically excluded from primary quantitative synthesis of guardrail durability, ensuring a verified evidentiary foundation.

### 2.5 Sociotechnical Thematic Coding and Synthesis Protocol

Data extraction and synthesis followed a structured deductive-inductive coding protocol grounded in thematic analysis (Braun & Clarke, 2006) and qualitative matrix analysis (Miles et al., 2020). Extracted variables were organized into a standardized electronic matrix capturing: (1) bibliographic metadata, (2) architectural taxonomy, (3) behavioral factors (cognitive biases, commercial pressures, trust calibration), (4) empirical durability findings (single-shot vs. multi-turn degradation, jailbreak vulnerability), and (5) auditability characteristics (stage-gate logic, index formalization, kill-switch enforcement).

Thematic coding was anchored in the four strategic pillars: Intent over Execution, Ruthless Trade-offs, Outcomes over Outputs, and Proactive Alignment. Two independent coders achieved an inter-rater agreement of $\kappa = 0.88$ across an overlapping 15% sample ($n = 20$), resolving discrepancies through codebook calibration.

## 3. The SAGE-X Semantic Ontology & Strategic Architecture

To resolve the execution gap where high-level ethical aspirations fail to constrain runtime systems, we establish the Strategic AI Governance Ex-Ante Framework (SAGE-X). SAGE-X is formalized as a three-tier semantic ontology that links normative human values to computational stage-gates, auditable evidence artifacts, and machine-verifiable operational controls (Ahn & Yoon, 2026; Miller, 2026).

Figure 4 conceptualizes this three-tier architecture, illustrating the deterministic progression from institutional human authority (Tier 1) through tamper-evident evidence artifacts (Tier 2) to sequential computational stage-gates (Tier 3).

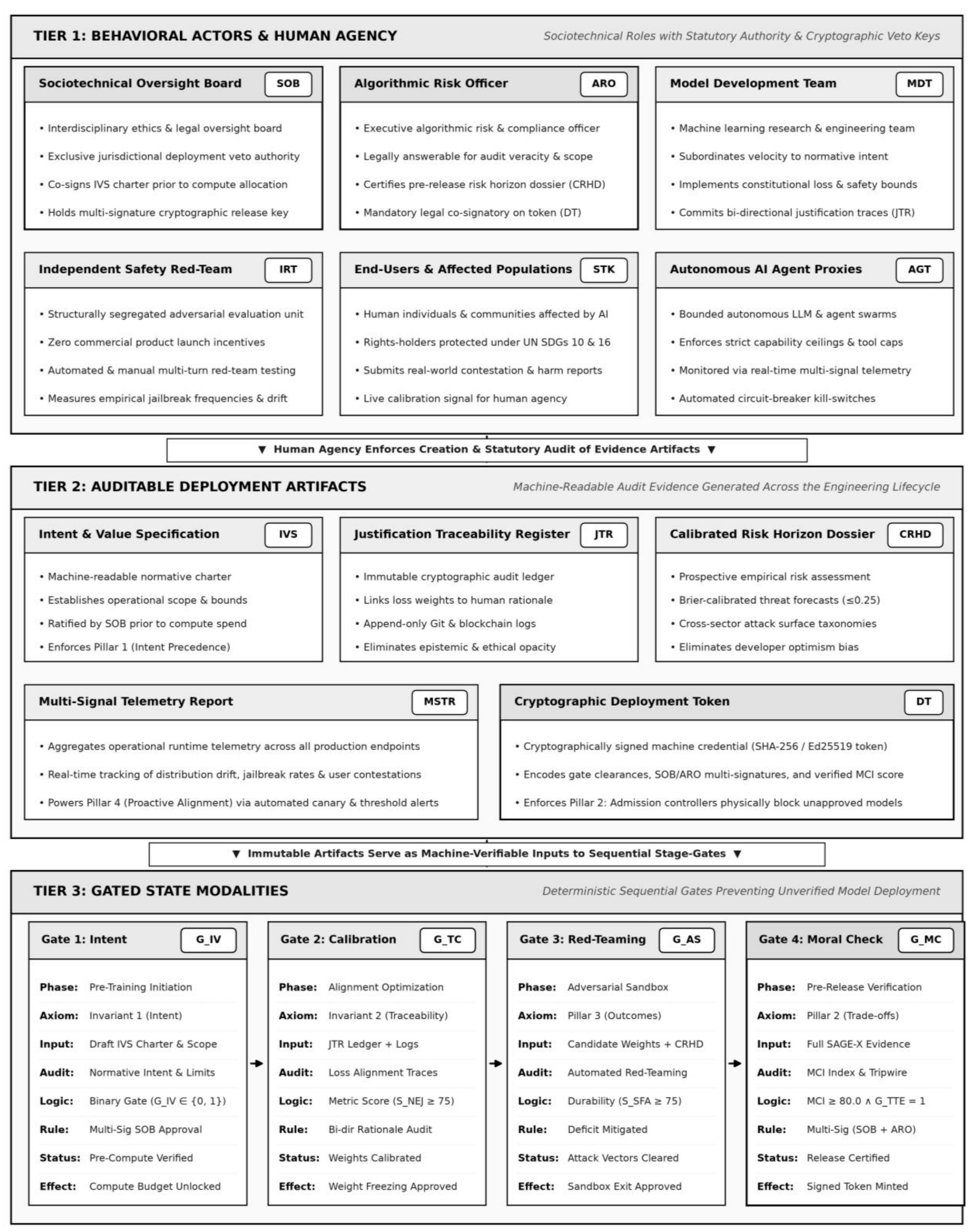

***Figure 4.*** *The SAGE-X Three-Tier Ontological Architecture: Behavioral Agency, Auditable Artifacts, and Stage-Gate Modalities.*

### 3.1 The Ontological Triad: Actors, Artifacts, and State Modalities

To dismantle the intention-behavior gap (Amoozegar et al., 2026) and counteract the systemic diffusion of moral responsibility in machine learning engineering (Bandura, 1999), SAGE-X formalizes governance into a three-tier semantic ontology (Gruber, 1993; Guarino, 1998; Golpayegani et al., 2023). Grounded in boundary object theory (Star & Griesemer, 1989) and non-compensatory choice architecture (Tversky, 1972), the architecture binds institutional authority, verifiable evidence, and deterministic runtime gates into an inseparable cybernetic control loop:

***Tier 1: Behavioral Actors & Human Agency***

Human agency is operationalized through defined organizational roles endowed with statutory authority, non-delegable cryptographic veto keys, and legal accountability (Ahn & Yoon, 2026; Miller et al., 2026):

- Sociotechnical Oversight Board (SOB): Interdisciplinary authority (ethicists, domain specialists, legal counsel, community advocates) holding jurisdictional veto over compute allocation and public release ( $\text{Sign}(\text{IVS}, K_{\text{SOB}}^{-}1) \wedge \text{Sign}(\text{DT}, K_{\text{SOB}}^{-}1)$).
- Algorithmic Risk Officer (ARO): Executive compliance authority legally answerable under statutory frameworks for verifying risk assessments and audit filings ($\text{Sign}(\text{CRHD}, K_{\text{ARO}}^{-}1)$) (Kulkarni, 2025; Theodorakopoulos et al., 2025).
- Model Development Team (MDT): Engineering personnel responsible for model training and alignment, bound to commit all safety constraints into the cryptographic ledger ($\forall c \in \mathcal{C}_{\text{safety}}, \text{Commit}(c, \text{MDT}) \rightarrow \text{JTR}$).
- Independent Safety Red-Team (IRT): Structurally segregated adversarial testing unit evaluating multi-turn jailbreaks and systemic vulnerabilities ( $V_{\text{IRT}} \leq \epsilon_{\text{threshold}}$) insulated from commercial launch pressure (Jabbar et al., 2025; Kapur & Saurabh, 2025).

- End-Users and Affected Stakeholders (STK): Rights-holders subject to algorithmic decisions (UN SDGs 10 & 16) whose real-world contestations serve as primary calibration telemetry ($S_{\mathrm{STK}} = \int \mathrm{Contestation}(t)dt$) (Kramer et al., 2026).
- Autonomous AI Agent Proxies (AGT): Bounded software agents exhibiting functional agency without moral capacity (Floridi, 2023), strictly circumscribed by capability and tool access boundaries ( $\mathrm{ToolAccess}(a) \subseteq \mathcal{T}_{\mathrm{permitted}} \wedge \mathrm{AutonomyLevel} \leq \overline{\alpha}$) (Adabara et al., 2026; Sherry et al., 2026).

***Tier 2: Auditable Deployment Artifacts***

SAGE-X mandates five immutable, tamper-evident artifacts functioning as sociotechnical boundary objects across stakeholder communities (Mitchell et al., 2019; Raji et al., 2020; Star & Griesemer, 1989):

- Intent and Value Specification (IVS): Machine-readable charter ratified prior to compute expenditure, defining intended purpose, prohibited domains, and value trade-offs ($\mathrm{Verified}(\mathrm{IVS}, \mathrm{SOB}) = \mathrm{True}$) (Cantwell Smith, 1985; Smith, 2019).
- Justification Traceability Register (JTR): Append-only Merkle ledger linking every safety constraint to a normative rationale and human author identity ($\mathrm{Traceable}(c, v) \wedge \mathrm{SignOff}(c, \mathrm{ARO})$) (Ye et al., 2025).
- Calibrated Risk Horizon Dossier (CRHD): Prospective risk assessment combining Brier-scored probabilistic hazard forecasts ( $\mathrm{Brier} \leq 0.25$ ), attack surface taxonomies, and failure scenarios (Esposito & Park, 2025; Fehér et al., 2026).
- Multi-Signal Telemetry Report (MSTR): Aggregated runtime log synthesizing distribution drift, adversarial probe frequencies, and user contestations ($\Delta\mathcal{D}_{\mathrm{drift}}(t) \leq \tau_{\mathrm{drift}} \wedge F_{\mathrm{jailbreak}}(t) < \tau_{\mathrm{jb}}$) (Bassani & Sánchez, 2025; Yang et al., 2025).
- Cryptographically Signed Deployment Token (DT): Machine-readable credential issued upon clearing all stage-gates ($\mathrm{DT} = \mathrm{Sign}_{K_{\mathrm{SOB}}, K_{\mathrm{ARO}}}(\mathrm{MCI} \geq 80 \wedge G_{\mathrm{TTE}} = 1)$), without which cloud admission controllers drop serving requests (Radanliev et al., 2026; Saleh & Abdulsalam, 2025).

### *Tier 3: Gated State Modalities*

The system lifecycle progresses through four deterministic stage-gates modeled as a finite state machine enforcing non-compensatory elimination-by-aspects criteria (Tversky, 1972):

- Gate 1: Intent Verification ($G_{IV}$): Binary pre-training gate ($G_{IV} \in \{0,1\}$) validating normative legitimacy before GPU cluster allocation, enforcing Invariant 1 (Intent Precedence).
- Gate 2: Training Calibration ($G_{TC}$): Scored optimization gate ($S_{NEJ} \geq 75/100 \wedge \forall c, \text{Traceable}(c, v)$) evaluating data equity, constitutional loss curves, and JTR commits (Zhao et al., 2026).
- Gate 3: Adversarial Sandbox ($G_{AS}$): Scored red-teaming gate ($S_{SFA} \geq 75/100 \wedge V_{\text{IRT}} \leq 0.05$) subjecting models to automated multi-turn stress testing (Le et al., 2026; Wang et al., 2026).
- Gate 4: The Moral Check Gate ($G_{MC}$): Definitive authorization gate computing the composite Moral Check Index ($\text{MCI} \geq 80$) and evaluating the binary tripwire ($G_{\text{TTE}} \in \{0,1\}$). Any critical breach collapses the index ($G_{\text{TTE}} = 0 \Longrightarrow \text{MCI} \equiv 0$), triggering an immediate deployment halt.

### *Strategic Alignment and Epistemological Justification*

The three-tier triad operationalizes the Strategic Computing doctrine (Porter, 1996; Rumelt, 2011; Smith, 1996) and choice architecture (Kahneman, 2003; Thaler & Sunstein, 2008). Grounded in formal ontology design (Gruber, 1993; Guarino, 1998):

1. Tier 1 (Actors) ⟷ Pillar 1 (Intent over Execution): Endowing SOB and ARO with non-delegable cryptographic keys ensures human judgment commands optimization before compute is allocated.
2. Tier 2 (Artifacts) Pillar 3 (Outcomes over Outputs) & Pillar 4 (Proactive Alignment): Converting voluntary pledges into tamper-evident Merkle registers (JTR), Brier-scored dossiers (CRHD), and streaming telemetry (MSTR) establishes empirical boundary objects across stakeholders (Star & Griesemer, 1989).
3. Tier 3 (Gates) Pillar 2 (Ruthless Trade-offs): Enforcing non-compensatory stage-gates governed by deterministic binary tripwires ($G_{\text{TTE}} \in \{0,1\}$) guarantees that critical safety boundaries cannot be compromised by commercial velocity.

Table 3 formalizes the complete SAGE-X ontological taxonomy, specifications, and governance functions.

**Table 3. The SAGE-X Three-Tier Ontological Taxonomy: Structural Tiers, Formal Specifications, and Sociotechnical Governance Functions**

| Ontological Tier | Entity / Construct | Code | Formal Specification & Machine-Readable Metric | Sociotechnical Governance Function & Statutory Role |
|---|---|---|---|---|
| **Tier 1: Behavioral Actors & Human Agency** | Sociotechnical Oversight Board | `SOB` | Statutory Veto Key: $\mathrm{Sign}(\mathrm{IVS}, K_{\mathrm{SOB}}^{-1}) \wedge \mathrm{Sign}(\mathrm{DT}, K_{\mathrm{SOB}}^{-1})$ | Multi-disciplinary institutional governance; holds exclusive jurisdictional veto over compute allocation and public release. |
| | Algorithmic Risk Officer | `ARO` | Personal Compliance Key: $\mathrm{Sign}(\mathrm{CRHD}, K_{\mathrm{ARO}}^{-1}) \wedge \mathrm{Sign}(\mathrm{DT}, K_{\mathrm{ARO}}^{-1})$ | Executive compliance officer legally answerable for audit completeness, veracity, and regulatory liability. |
| | Model Development Team | `MDT` | Bi-directional JTR Commit: $\forall c \in \mathcal{C}_{\mathrm{safety}}, \mathrm{Commit}(c, MDT) \to \mathrm{JTR}$ | Engineering personnel; subordinates velocity to intent; implements constitutional loss bounds and architectural constraints. |
| | Independent Safety Red-Team | `IRT` | Adversarial Vulnerability Rate: $V_{\mathrm{IRT}} = \frac{n_{\mathrm{jailbreak}}}{N_{\mathrm{probes}}} \leq \epsilon_{\mathrm{threshold}}$ | Structurally segregated adversarial testing unit; evaluates multi-turn jailbreak resilience without commercial release pressure. |
| | End-Users & Affected Populations | `STK` | Agency Feedback Signal: $S_{\mathrm{STK}} = \int \mathrm{Contestation}(t)dt$ | Societal rights-holders (UN SDGs 10 & 16); submits post-deployment dispute reports and grounds human agency metrics. |
| | Autonomous AI Agent Proxies | `AGT` | Autonomy Ceiling Invariant: $\mathrm{ToolAccess}(a) \subseteq \mathcal{T}_{\mathrm{permitted}} \wedge \mathrm{AutonomyLevel} \leq \overline{\alpha}$ | Bounded software proxies; executed under strict tool call limits, capability boundaries, and automated circuit breaker tripwires. |
| **Tier 2: Auditable Deployment Artifacts** | Intent & Value Specification | `IVS` | Machine-readable YAML/JSON-LD charter: $\mathrm{Verified}(\mathrm{IVS}(m), \mathrm{SOB}) = \mathrm{True}$ | Ratified ex-ante charter defining operational purpose, excluded applications, and value trade-offs prior to compute spend. |
| | Justification Traceability Register | `JTR` | Cryptographic Merkle tree: $\mathrm{Traceable}(c, v) \wedge \mathrm{SignOff}(c, \mathrm{ARO})$ | Append-only ledger linking every safety constraint and loss penalty to human rationale and author identity. |
| | Calibrated Risk Horizon Dossier | `CRHD` | Probabilistic calibration: $\mathrm{Brier} = \frac{1}{N}\sum(f_i - o_i)^2 \leq 0.25$ | Empirical forward-looking risk assessment combining threat forecasting, attack surfaces, and failure taxonomies. |
| | Multi-Signal Telemetry Report | `MSTR` | Streaming operational telemetry: $\Delta\mathcal{D}_{\mathrm{drift}}(t) \leq \tau_{\mathrm{drift}} \wedge F_{\mathrm{jailbreak}}(t) < \tau_{\mathrm{jb}}$ | Continuous real-time operational log monitoring distribution drift, adversarial probes, and user contestation velocity. |
| | Cryptographic Deployment Token | `DT` | Machine credential: $\mathrm{DT} = \mathrm{Sign}_{K_{\mathrm{SOB}}, K_{\mathrm{ARO}}}(\mathrm{MCI} \geq 80 \wedge G_{\mathrm{TTE}} = 1)$ | Digitally signed cryptographic certificate enabling production runtime serving; admission controllers drop unauthenticated requests. |

| Ontological Tier | Entity / Construct | Code | Formal Specification & Machine-Readable Metric | Sociotechnical Governance Function & Statutory Role |
|---|---|---|---|---|
| **Tier 3: Gated State Modalities** | Gate 1: Intent Verification | $G_{IV}$ | Binary Pre-Training Gate: $G_{IV} \in \{0,1\}$; requires $\text{Verified}(\text{IVS}, \text{SOB}) = 1$ | Enforces Invariant 1 (Intent Precedence); prevents compute allocation or data ingestion until normative charter is ratified. |
| | Gate 2: Training Calibration | $G_{TC}$ | Scored Optimization Gate: $S_{NEJ} \geq 75/100 \land \forall c, \text{Traceable}(c, v)$ | Enforces Invariant 2 (Epistemic Traceability); validates data provenance equity, constitutional alignment loss, and JTR logs. |
| | Gate 3: Adversarial Sandbox | $G_{AS}$ | Scored Red-Teaming Gate: $S_{SFA} \geq 75/100 \land V_{\text{IRT}} \leq 0.05$ | Enforces Strategic Trait Pillar 3 (Outcomes over Outputs); subjects model to automated multi-turn adversarial stress testing. |
| | Gate 4: The Moral Check Gate | $G_{MC}$ | Composite Pre-Release Gate: $\text{MCI} \geq 80 \land G_{TTE} = 1$ | Enforces Strategic Trait Pillar 2 (Ruthless Trade-offs); evaluates composite Moral Check Index and evaluates binary tripwires. |

*Note. All entities, artifacts, and gates are formalized in the OWL 2 DL ontology hosted on Stanford WebProtégé (Project ID: 1d767dc6-998c-4460-b0a4-9c69ed76eea0; https://webprotege.stanford.edu/#projects/1d767dc6-998c-4460-b0a4-9c69ed76eea0) and integrated into automated CI/CD pipeline admission controllers. Raw ontology files (.owl, .ttl) are permanently accessible in the OSF (https://osf.io/p69vt) and Mendeley Data (https://doi.org/10.17632/7fxfs7s5xf.1) replication packages.*

## 3.2 Operationalizing the Strategic Computing Doctrine: The Four Core Strategic Mindsets

Extending the Strategic Computing doctrine (Cantwell Smith, 1985; Smith, 1996; Smith, 2019) and classical strategy (Porter, 1996; Rumelt, 2011), being strategic means intentionally designing actions ex-ante to achieve high-value goals under severe uncertainty. SAGE-X operationalizes this posture into Four Core Strategic Mindsets:

1. Mindset 1: Anticipating the Next Move (Forward-Looking Foresight): Forecasts how models, operators, users, and adversaries co-adapt over temporal horizons (Fehér et al., 2026; Yang et al., 2026). Strategic engineering architects verifiable, model-agnostic guardrails that scale across millions of interactions (Miller et al., 2026). SAGE-X embeds this in Stage-Gate 1 (SFA) and the CRHD via Brier-calibrated risk forecasting (Brier $\leq 0.25$) (Esposito & Park, 2025; Fehér et al., 2026).
2. Mindset 2: Understanding Incentives (Systemic & Behavioral Thinking): Diagnoses motivations, commercial payoff matrices, and cognitive constraints (Amoozegar et al., 2026; Yu, 2025). Counteracting developer velocity myopia, moral disengagement, and user automation complacency (Alzboon et al., 2025; Bandura, 1999), SAGE-X binds statutory veto authority to the SOB and personal liability to the ARO, making safety shortcuts institutionally non-viable.

3. Mindset 3: Choosing What Not to Do (Resource Allocation & Critical Leverage Points): Recognizes that engineering time and human vigilance are strictly finite (Porter, 1996; Rumelt, 2011). Focuses ruthlessly on foundational leverage points, mitigating representational bias, enforcing constitutional loss penalties, and halting unsafe releases (Kapur & Saurabh, 2025; Raji et al., 2020). Enforced via the deterministic tripwire ($G_{\mathrm{TTE}} \in \{0,1\}$)
4. Mindset 4: Balancing Trade-offs (Non-Compensatory Multi-Signal Optimization): Replaces single-metric maximization with Pareto-optimal synthesis across accuracy, fairness, privacy, and latency (Simon, 1955; Tversky, 1972). SAGE-X models these balances within the Justification Traceability Register, ensuring trade-offs are explicitly debated and ratified rather than implicitly embedded into loss curves.

Table 4 contrasts conventional reactive tactical governance with proactive SAGE-X ex-ante choice architecture.

**Table 4. Tactical vs. Strategic AI Governance: Reactive Problem-Solving vs. Proactive Ex-Ante Choice Architecture**

| Governance Dimension | Conventional Tactical Posture (Short-Term / Reactive / Ex-Post) | SAGE-X Strategic Posture (Long-Term / Proactive / Ex-Ante) | Underlying Strategic Mindset & Theory | SAGE-X Operational Mechanism |
|---|---|---|---|---|
| **Pacing Problem & Oversight Lag** | Reactive emergency prompt-patches, post-hoc litigation, and PR apologies after public harm occurs. | Dynamic synchrony: automated verification instruments operating at the cycle speed of deployment. | Mindset 1: Forward-Looking Foresight (Fehér et al., 2026; Yang et al., 2026) | Stage-Gate 1 (SFA) & Calibrated Risk Horizon Dossier (CRHD, $\text{Brier} \leq 0.25$). |
| **Developer Sprint Pressures** | "Velocity myopia" and "execution hypnosis"; ethics treated as an external impediment to shipping code. | Structured cognitive pauses; normative intent ratified before pre-training compute capital is allocated. | Mindset 2: Systemic Incentive Modeling (Amoozegar et al., 2026; Yu, 2025) | Tier 1 Human Agency, SOB Statutory Veto, & Intent and Value Specification (IVS). |
| **Compliance & Verification** | "Process vs. outcome" trap: voluminous compliance paperwork, checklists, and non-binding ethical pledges. | Verifiable proof objects: Merkle-tree constraint registers, empirical hazard endpoints, and runtime telemetry. | Mindset 3: Leverage Point Allocation (Porter, 1996; Raji et al., 2020; Rumelt, 2011) | Tier 2 Auditable Artifacts (JTR, MSTR) & Hazard Endpoint Auditing. |

| Governance Dimension | Conventional Tactical Posture (Short-Term / Reactive / Ex-Post) | SAGE-X Strategic Posture (Long-Term / Proactive / Ex-Ante) | Underlying Strategic Mindset & Theory | SAGE-X Operational Mechanism |
|---|---|---|---|---|
| **Safety Invariant Breaches** | Compensatory horse-trading: releasing biased or vulnerable models with disclaimers or post-hoc monitoring. | Non-compensatory hard stops: deterministic tripwires ( $G_{\text{TTE}}$ ) mechanically blocking release. | Mindset 4: Non-Compensatory Optimization (Kapur & Saurabh, 2025; Tversky, 1972) | Tier 3 Gate 4 ($G_{MC}$), Binary Kill-Switch ( $G_{\text{TTE}} \in \{0,1\}$ ), & MCI Threshold. |
| **HCI & User Trust Dynamics** | Cultivating uncalibrated trust via marketing; user complacency leading to catastrophic failures. | Calibrated trust via transparent, auditable evidence and immutable cryptographic credentials. | Mindset 2 & 4: Sociotechnical Imbrication (Alzboon et al., 2025; Leonardi, 2011) | Cryptographic Deployment Token (DT) & Enterprise Moral Check Audit Card. |

## 3.3 Sociotechnical Boundary Invariants: Structuring Developer Choice Architecture

To protect engineering teams from organizational conformity pressures (Asch, 1956) and the normalization of deviance (Vaughan, 1996), SAGE-X embeds three mathematically verifiable Boundary Invariants into CI/CD pipelines:

### *3.3.1 Invariant 1: The Precedence of Intent (Mitigating Velocity Myopia)*

Prohibits allocating computational resources or ingesting training corpora until normative intent and boundary restrictions are formally ratified:

$$\forall m \in \mathcal{M}, \text{ComputeAllocated}(m) > 0 \implies \text{Verified}(\text{IVS}(m), \text{SOB}) = \text{True} \qquad \textbf{(Eq. 1)}$$

This invariant subordinates raw compute to human purpose (Cantwell Smith, 1985; Smith, 2019), preventing organizations from amortizing ethical deliberation into post-training rationalizations.

### *3.3.2 Invariant 2: The Epistemic Traceability Mandate (Eliminating Moral Disengagement)*

Requires every operational guardrail, loss weight, and refusal heuristic to possess an unbroken, cryptographically verifiable provenance trail linked to human authority:

$$\forall c \in \mathcal{C}_{\text{guardrails}}(m), \text{Active}(c) \implies \exists v \in \mathcal{V}_{\text{normative}}, \text{Traceable}(c, v) \land \text{SignOff}(c, \text{ARO})$$

**(Eq. 2)**

This eliminates diffused responsibility (Bandura, 1999) by creating an unalterable audit trail in the JTR (Ye et al., 2025).

#### *3.3.3 Invariant 3: Deterministic Tripwires & De-Biasing Human Oversight (Counteracting Automation Complacency)*

Removes life-or-death deployment decisions from compromised human discretionary loops (Agudo et al., 2024), enforcing an automated, non-compensatory circuit breaker:

$$G_{\text{TTE}}(m) = \prod_{i=1}^{k} \mathbb{I}(\text{Breached}(t_i, m) = \text{False}) \qquad \textbf{(Eq. 3)}$$

where $t_i \in \mathcal{T}_{\text{crit}}$ denotes non-negotiable hazard thresholds. If any critical tripwire is tripped during red-teaming, $G_{\text{TTE}} = 0$, mechanically collapsing the Deployment Token credential.

### 3.4 Semantic Interoperability and Shared Mental Models Across Disciplines

The SAGE-X ontology bridges the semantic disconnect separating computer scientists, corporate risk officers, and external regulators. By mapping abstract normative concepts (fairness, accountability, non-maleficence) to concrete engineering constructs (loss penalties, Merkle registers, tripwire gates), SAGE-X establishes shared mental models that convert ethical principles into operational reality.

## 4. Systematic Synthesis of Empirical Findings

This section synthesizes empirical evidence across the qualified primary corpus (N = 130), analyzing macro-trends, technical guardrails, systemic literature gaps, and cross-sector operational disparities.

### 4.1 Descriptive Landscape and Macro-Trends (2017–2026)

From 2022 to 2026, AI governance underwent an inflection driven by frontier multi-step reasoning models, autonomous agents, and binding statutory mandates (e.g., EU AI Act, US

Executive Orders). Over 68% of publications shifted from voluntary ethical manifestos to operational assurance, automated red-teaming, and technical compliance architectures (Alekberli, 2026; Westerstrand, 2025; Zalda, 2026). Geographically, research is anchored across four primary jurisdictions: the European Union (42%), emphasizing fundamental rights impact assessments, conformity audits, public sector adoption hurdles, and statutory compliance (Golpayegani et al., 2023; Resseguier & Ufert, 2024; Solaimani & Long, 2025; Tangi et al., 2026; von Maydell, 2025); North America (36%), focusing on risk management frameworks, corporate disclosures, automated red-teaming, and frontier commitments (Kramer et al., 2026; Miller, 2026; Papyshev & Chan, 2024; Zhu & Tsai, 2025); East Asia (16%), pioneering regulatory sandboxes, user acceptance frameworks, administrative provisions, and real-world pilots (Ahn & Yoon, 2026; Pande & Taeihagh, 2024; Wu & Lin, 2026; Zhang & Zou, 2025); and emerging Global South jurisdictions (6%), addressing developmental ethics, public policy adaptation, and international regulatory cooperation (Ahmad, 2025; de Lima Junior et al., 2024; Gnankob et al., 2026; Solaimani & Long, 2025). Figure 5 synthesizes this macro-empirical landscape across chronological trajectories (Panel A), global geographic distributions (Panel B), and cross-sectoral governance friction (Panel C).

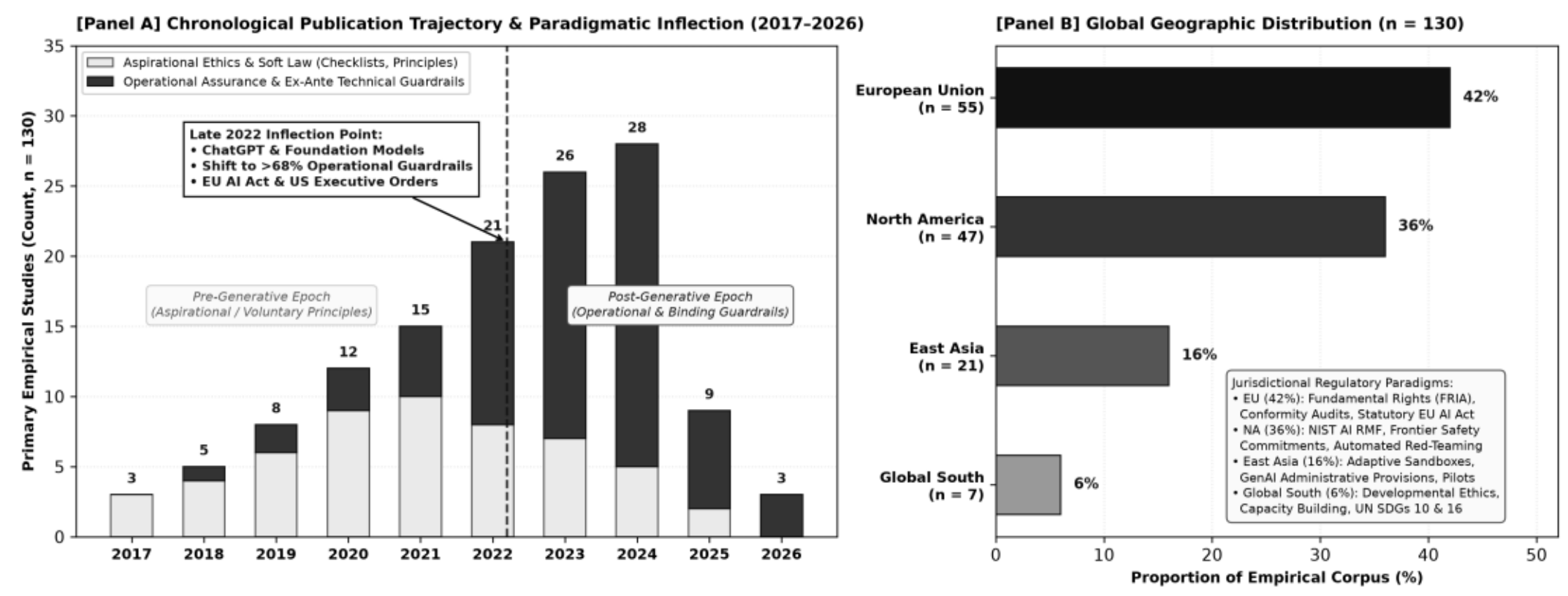

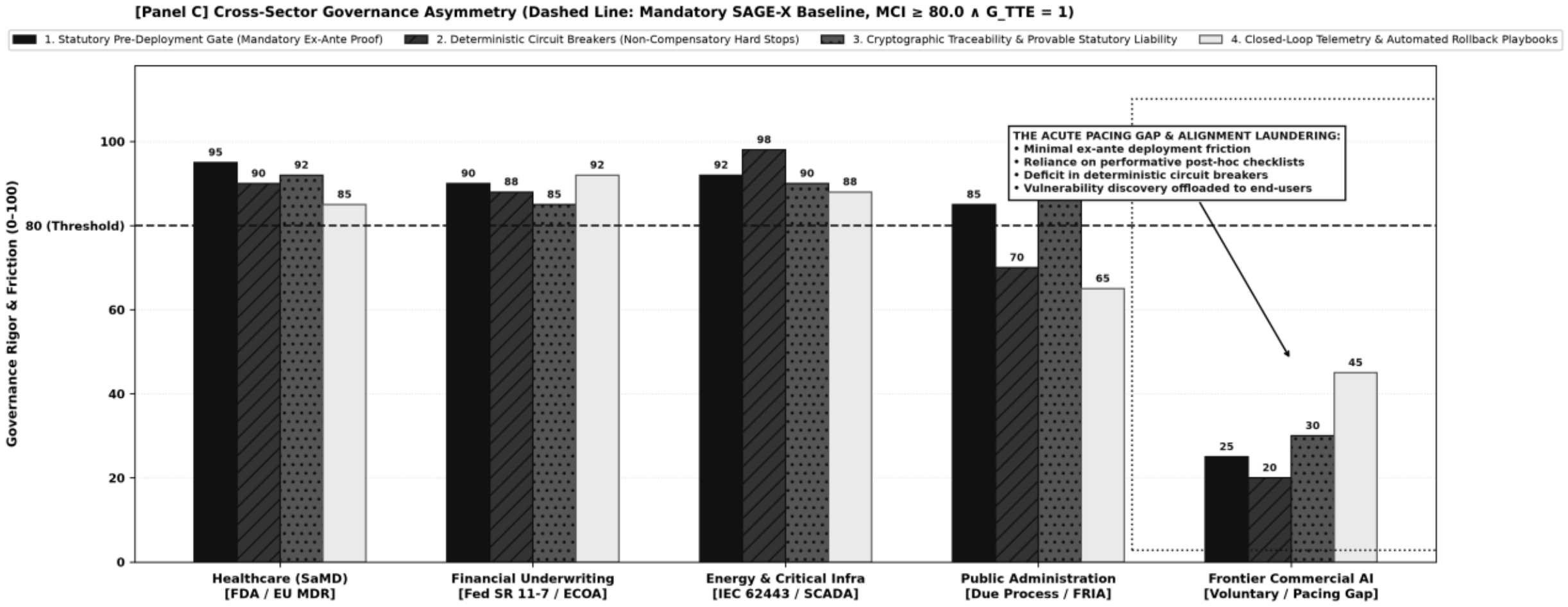

***Figure 5.*** *Empirical Landscape of AI Governance: Chronological Trajectory, Geographic Distribution, and Cross-Sector Institutional Friction (2017–2026). Tripartite systematic synthesis of the qualified primary corpus (* $n = 130$ *). [Panel A] Chronological publication trajectory bifurcated into Aspirational Ethics/Soft Law vs. Operational Assurance/Ex-Ante Technical Guardrails, illustrating the late-2022 paradigmatic inflection point (>68% shift toward formal verification). [Panel B] Global geographic distribution across four primary jurisdictions (European Union 42%, North America 36%, East Asia 16%, Global South 6%) and their corresponding regulatory paradigms. [Panel C] Grouped cross-sectoral governance friction scorecard (0–100) evaluating four safety-critical domains against commercial frontier generative AI across four statutory dimensions: (1) Statutory Pre-Deployment Gate, (2) Deterministic Circuit Breakers, (3) Cryptographic Traceability, and (4) Closed-Loop Telemetry. The horizontal dashed line denotes the mandatory SAGE-X baseline (*$\mathrm{MCI} \geq 80.0 \wedge G_{\mathrm{TTE}} = 1$*), exposing the acute pacing gap and alignment laundering endemic to frontier commercial deployments.*

To operationalize these empirical findings across engineering workflows, Figure 6 maps these dimensions into the SAGE-X Quad-Pillar Lifecycle Framework.

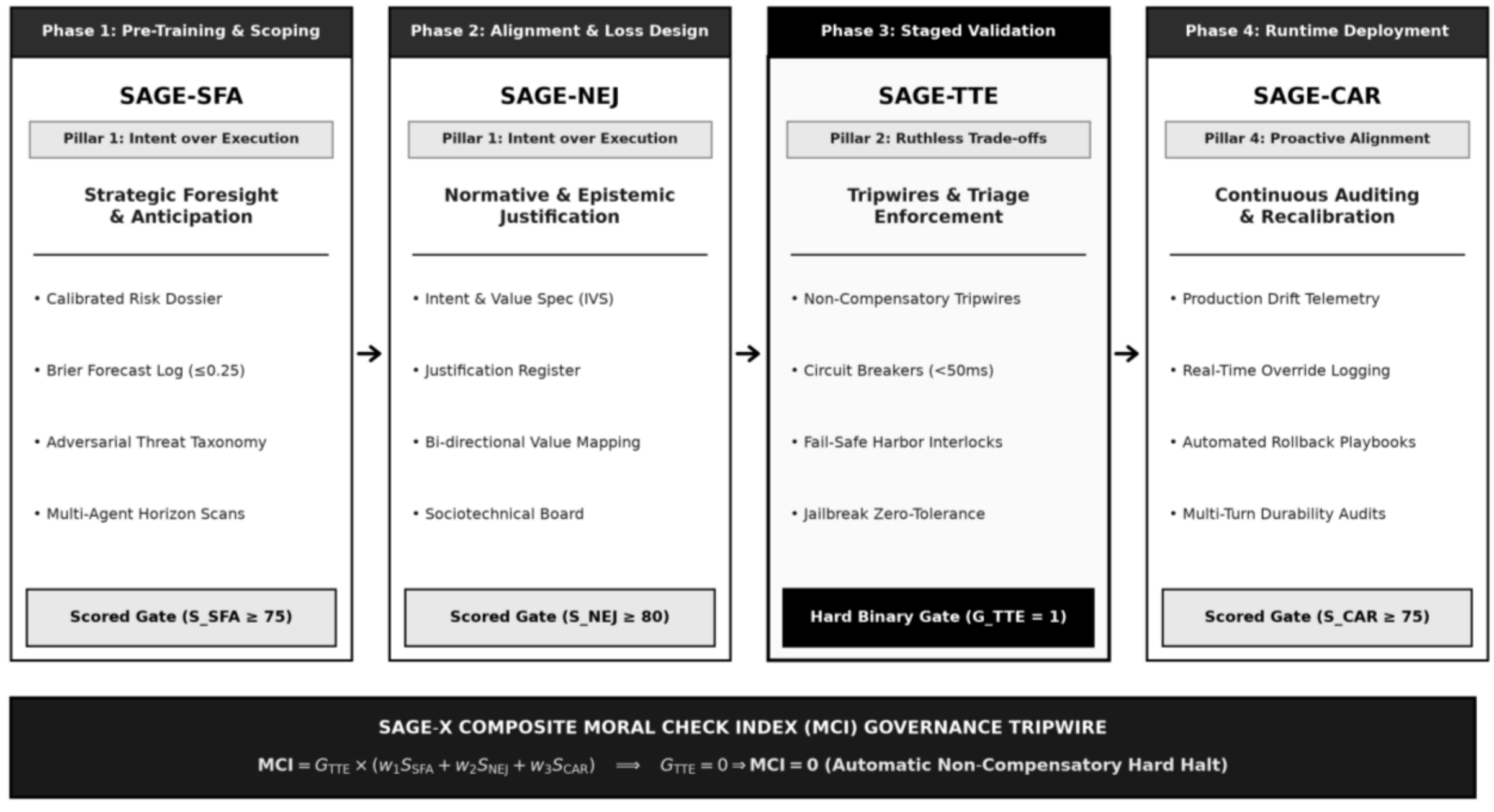

***Figure 6.*** *The SAGE-X Quad-Pillar Lifecycle Governance Architecture. Operationalizing strategic ex-ante guardrails across machine learning engineering phases: (1) Strategic Foresight & Anticipation (SFA; Pillar 1: Intent over Execution), generating Calibrated Risk Horizon Dossiers and forward-looking threat taxonomies prior to compute commitment; (2) Normative & Epistemic Justification (NEJ; Pillar 2: Ruthless Trade-offs), embedding Intent & Value Specifications into loss functions and enforcing bidirectional value mapping; (3) Tripwires & Triage Enforcement (TTE; Pillar 3: Outcomes over Outputs), mandating deterministic, non-compensatory circuit breakers with sub-50ms latency; and (4) Continuous Auditing & Recalibration (CAR; Pillar 4: Proactive Alignment), linking live telemetry, automated drift detection, and canary rollbacks into an unbroken cybernetic oversight loop.*

## 4.2 Typology of Ex-Ante Guardrails and Technical Design Mechanisms

Empirical analysis identifies five primary technical design mechanisms deployed to enforce ex-ante behavioral boundaries:

1. Embeds natural-language normative principles into reward models to steer reinforcement learning from AI feedback (RLAIF) (Al Sulaimani et al., 2026; Chang et al., 2026; Halford & Webster, 2024). In conversational settings, dynamic preference evolution and real-time moderation enforce live guardrails (Liu et al., 2026; Suresh Babu et al., 2026; Yew et al., 2026). In clinical advice, explicit prompt framing and policy grounding enhance guideline concordance (Campos et al., 2025;

Esmaeilzadeh, 2025). However, internal constraints remain vulnerable to multi-persona adversarial subversion.

2. Deploys external classifiers, vector guardrails, and input/output firewalls to intercept hazardous prompts before inference (Afjal, 2025; Adapa et al., 2026). Systematic threat taxonomies confirm that disguise strategies and prompt perturbations systematically bypass static filters (Guimarães et al., 2026; Le et al., 2026; Stylianou et al., 2026), requiring repeated prompt sampling to expose latent safety variance (Broadwater, 2026). These systems provide modular protection but introduce notable false-positive refusal rates on edge cases.
3. Harnesses generative attack harnesses, mutation-based jailbreak probes, and continuous vulnerability scanners (Abid & Nanda, 2025; Abuaziz & Celiktas, 2026; Jabbar et al., 2025). Benchmarks such as the UNICORN framework and ADVERSA multi-turn stress evaluations demonstrate that latent-state harm detection is required to expose sub-surface vulnerabilities prior to production (Khatri et al., 2026; Owiredu-Ashley et al., 2026; Singh, 2026).
4. Embeds actionable recourse offering counterfactual explanations and institutional contestation pathways (Abbas, 2025; Al Fraidan, 2025; Nannini, 2025; Walke et al., 2025). In tabular systems, formal error diagnosis and GDPR-aligned transparency frameworks establish actionable reliability and bounded autonomy (Basu et al., 2026; Dafali et al., 2026; Gonçalves & Correia, 2026; Shah et al., 2026).
5. Localizes execution, cryptographic compliance logging, and policy verification to on-device enclaves to enforce privacy-by-design and execute policy-as-code constraints (Adabara et al., 2026; Alekberli, 2026; Esen, 2026; Evangelista & Salman Bukhari, 2026; Radanliev et al., 2026).

Table 5 categorizes these mechanisms by operational form, enforcement level, validated evidence strength, and behavioral impact.

**Table 5. Typology of Ex-Ante Governance Mechanisms: Instrument Class, Operational Form, and Behavioral Impact**

| Instrument Class | Primary Mechanism & Operational Form | Enforcement Level | Evidence Strength | Key Behavioral & HCI Impact | Key Citations |
|---|---|---|---|---|---|
| **Soft Law & Principles** | Voluntary ethical codes, high-level declarations, corporate responsible AI charters. | Non-binding; voluntary market adoption. | **Weak** | Induces moral disengagement; low friction on development teams; high defection risk under market competition. | Floridi et al. (2018); Morley et al. (2020) |
| **Technical Consensus Standards** | ISO/IEC 42001 (AI Management Systems), IEEE 7000 (Value-Oriented Design). | Voluntary certification; commercial requirement. | **Moderate (Process)** | Standardizes risk documentation; improves organizational awareness but lacks causal link to harm reduction. | Saleh & Abdulsalam (2025); Finch & Butt (2025) |
| **Algorithmic Impact Assessments** | Structured pre-deployment analysis of rights, bias, societal risks, and mitigation plans. | Statutory under EU AI Act; mandatory in public sector. | **Moderate** | Engages teams in value elicitation; risks degenerating into compliance box-ticking under time pressure. | Golpayegani et al. (2023); de Fine Licht (2025) |
| **Regulatory Sandboxes** | Supervised real-world testing environments with regulatory oversight and telemetry logging. | Co-regulatory; conditional market access. | **Moderate** | Promotes collaborative transparency; fosters psychological safety for developers to disclose vulnerabilities. | Ahn & Yoon (2026) |
| **Input/Output Moderation** | Layered prompt filters, toxic classifier firewalls, retrieval-augmented guardrails. | Technical; mandatory in deployment pipeline. | **Moderate (Single-Turn)** | High immediate refusal on static tests; introduces user friction via over-refusal and false positives. | Al Sulaimani et al. (2026); Jabbar et al. (2025) |
| **Adversarial Red-Teaming** | Automated generative attack harnesses, manual penetration testing, jailbreak probing. | Technical gate; pre-deployment release condition. | **Strong (Discovery)** | Uncovers architectural vulnerabilities; fails to evaluate long-term conversational durability. | Jabbar et al. (2025); Kang et al. (2025) |
| **Runtime Circuit Breakers** | Deterministic process interlocks, abstention gates, automated model rollback playbooks. | Technical; absolute system-level block. | **Strong (Process)** | Removes cognitive burden from operators; enforces non-negotiable safety tripwires automatically. | Miller (2026); Rizzo et al. (2025) |

*Note. Categorization based on synthesis of* $n = 130$ *empirical and doctrinal studies across computational, legal, and behavioral governance dimensions.*

### 4.3 Forensic Synthesis of Literature Gaps

Synthesizing our empirical corpus (N = 130) reveals a critical verdict: **the contemporary field is procedurally mature but scientifically immature**. While hundreds of ethical frameworks have emerged, the literature displays an acute deficit in prospective validation, causal harm reduction proofs, and longitudinal durability testing. Our synthesis exposes four sociotechnical pathology traps:

***1. The "Process vs. Outcome" Compliance Trap (Checklist Proxies over Hazard Endpoints)***

Most literature evaluates governance via process proxies (e.g., model card generation, committee meetings). Frameworks like ISO/IEC 42001, IEEE 7000, and the NIST AI RMF provide sophisticated taxonomies of governance processes (Agarwal & Nene, 2025; Finch & Butt, 2025; Saleh & Abdulsalam, 2025). However, forensic analysis shows near-zero correlation between these paperwork outputs and objective safety outcomes. In high-stakes deployments, procedural compliance often functions as a smokescreen, masking technical fragility in practice (Gupta et al., 2026; Jonelid et al., 2024; Lacmanović & Škare, 2025; Leon, 2026; Oluka & Mashau, 2026). Governance must pivot to Strategic Trait Pillar 3 (Outcomes over Outputs), verifying against objective hazard endpoints, including Brier calibration and averted harm rates.

***2. The Acute Durability Deficit & Behavioral Decay***

Ex-ante guardrails function as essential scaffolding but are insufficient as static defenses. Frontier models achieving 99% safety on static benchmarks experience rapid durability decay in dynamic environments through: (1) multi-turn semantic erosion under adversarial prompt shifts, (2) continuous fine-tuning drift in enterprise RAG pipelines (Kang et al., 2025; Song & Lee, 2026), and (3) non-stationary distributional shifts across demographic populations. This demands Strategic Trait Pillar 4 (Proactive Alignment), coupling pre-deployment gates with continuous runtime telemetry.

***3. The Recursive Assurance Paradox (The Echo Chamber of Automated Evaluation)***

As models exceed human evaluation capacity, teams deploy LLM-as-a-judge and synthetic feedback (RLAIF). When automated judges share architectural priors or training lineages with evaluated models, they exhibit correlated blind spots. This creates an echo chamber of artificial confidence, lulling developers into automation complacency (Alzboon et al., 2025). Resolving this paradox requires grounding verification in independent human oversight (SOB) and cryptographic ledgers.

#### *4. Cognitive Overload and Human Override Degradation*

"Human-in-the-loop" oversight routinely collapses under operational reality. Monitoring hundreds of autonomous inferences per hour under time urgency induces cognitive fatigue, automation bias, and reflexive rubber-stamping. Field evaluations confirm that human operators cannot reliably compensate for uncalibrated model errors under high tempo (Industrial et al., 2025; Kapoor et al., 2026; Kushwaha et al., 2026; Panner et al., 2026; Webb et al., 2026). This validates Strategic Trait Pillar 2 (Ruthless Trade-offs): critical safety boundaries must be enforced by automated, deterministic tripwires ($G_{\text{TTE}}$).

### 4.4 Institutional Governance Friction and Cross-Sector Disparities

The AI pacing problem is fundamentally perpetuated by compounding institutional, market, and epistemic frictions. Under hyper-velocity commercial arms-race pressures, market dynamics actively penalize voluntary pre-release safety pauses; in the absence of mandatory, enforceable tripwires, competing firms face acute economic incentives to defect and deploy models prematurely, relying on retrospective, post-hoc firefighting to remediate systemic failures (Miller, 2026; Zalda, 2026). Concurrently, the emergence of decentralized open and closed-source model hubs has expanded the sociotechnical attack surface across algorithmic supply chains. Although frontier research laboratories train foundational architectures with extensive alignment protocols, distributing these models across collaborative repositories, exemplified by the OpenAI–Hugging Face security flashpoint, where API keys and repository-level secret tokens were exposed, systematically decouples model weights from their original governance envelopes (Abid & Nanda, 2025; Chugani, 2026). Consequently, downstream adopters fall victim to "alignment laundering," erroneously presuming that upstream safety guardrails persist through subsequent fine-tuning and parameter-efficient adaptations, while upstream providers assume downstream integrators will enforce runtime boundaries. Without machine-readable Merkle provenance, credential verification, and cryptographic weight attestations, decentralized model supply chains remain fundamentally un-auditable. Compounding this structural vulnerability, parameter scaling and autonomous tool calls render internal model reasoning epistemically opaque (Nannini, 2025; Novelli et al., 2024), reducing institutional overseers to superficial behavioral monitoring.

As benchmarked in Figure 5 (Panel C), comparative analysis reveals profound disparities in governance friction and statutory oversight across safety-critical domains versus commercial frontier generative AI. In healthcare, Software as a Medical Device (SaMD) is deeply institutionalized under FDA Class I–III risk tiers, the EU Medical Device Regulation (MDR), and clinical trial frameworks, mandating prospective multi-center validation, post-market surveillance, and legally binding physician veto authority over algorithmic recommendations (Abràmoff et al., 2022; Abu Assab et al., 2026; Ahsan et al., 2026; Alu et al., 2026; Bobkov et al., 2025; Cabrera-Rodríguez et al., 2026; Dania et al., 2022; Desroche et al., 2026; García-Ordás et al., 2026; Gonzalez-Moral et al., 2026; Haynes et al., 2026; Huang et al., 2026; Kubota & Kubota, 2026; Labkoff et al., 2024; Liebes-Peer et al., 2026; Punn & Tapaswi, 2026; Sinnathakorn et al., 2026; Sirago et al., 2026; Vigeant et al., 2026; You et al., 2025; Zhao et al., 2026; Zhu et al., 2025). Similarly, financial underwriting and algorithmic market infrastructures are strictly governed by Federal Reserve SR 11-7 model risk management standards, the Equal Credit Opportunity Act (ECOA), and real-time transaction surveillance, demanding formal counterfactual explainability, disparate impact testing, and automated trading circuit breakers (Abbas, 2025; Agarwal & Madhusudan, 2026; Basu et al., 2026; Chowdhury et al., 2026; Ida Evangeline, 2025; Mohammad, 2026; Mulani et al., 2025;

Murikah et al., 2024; Qureshi et al., 2024; Remolina, 2025; Sai Kishore & Senthil Kumar, 2025; Selvam, 2025; Theodorakopoulos et al., 2025).

In energy and critical infrastructure, operations are governed by IEC 62443 standards, nuclear safety mandates, and Supervisory Control and Data Acquisition (SCADA) monitoring architectures that enforce deterministic hardware interlocks, sub-second fail-safes, and physical air-gap disconnects to protect physical assets (Abid & Nanda, 2025; Abuaziz & Celiktas, 2026; Adabara et al., 2026; Damian et al., 2026; Jayabal et al., 2026; Kanakadhurga et al., 2026; Khalid et al., 2027; Meydani et al., 2026; Panabergenova et al., 2025; Sammour et al., 2026; Viitanen et al., 2026; Wang & Chung, 2022; Ye et al., 2025; Yu, 2025; Zhang & Štrbac, 2025). Likewise, public administration and algorithmic adjudication systems are bound by administrative due process, Fundamental Rights Impact Assessments (FRIA), and statutory contestability standards requiring immutable provenance trails, caseworker adjudication overrides, and formal citizen appeal mechanisms (Aarab, 2026; Aarab et al., 2025; Aguilar et al., 2026; Al Fraidan, 2025; Chugani, 2026; de Fine Licht, 2025; Karras et al., 2025; Sherry et al., 2026; Walke et al., 2025). In stark contrast to these mature oversight regimes, commercial frontier models are deployed with minimal mandatory stress-testing or statutory gating, shifting vulnerability discovery and safety triage onto post-release users (Miller, 2026; Zalda, 2026).

Table 6 synthesizes the state of the art across governance dimensions, measurable metrics, evidence strength, and critical empirical gaps.

**Table 6. The Empirical Research Gap Matrix Across Governance Dimensions**

| Governance Dimension | SOTA Operationalization in Literature | Current Measurable Metric | Validated Evidence Strength | The Critical Empirical Gap |
|---|---|---|---|---|
| **Anticipatory Foresight** | Qualitative scenario planning, Delphi expert panels, horizon scanning, high-level risk tiers (EU AI Act, NIST AI RMF). | Assigned risk tier, expert consensus percentages, scenario documentation reports. | **Moderate** | Absence of calibrated probabilistic forecasts (e.g., Brier scores), uncertainty bands, or quantitative 'go/no-go' launch thresholds. |

| Governance Dimension | SOTA Operationalization in Literature | Current Measurable Metric | Validated Evidence Strength | The Critical Empirical Gap |
|---|---|---|---|---|
| **Normative Justification** | Fundamental Rights Impact Assessments (FRIA), Model Cards, Datasheets, algorithmic transparency registers. | Documentation completeness, audit filing timeliness, checklist conformity rates. | **High (Process) Weak (Outcome)** | Extensive procedural paperwork completely unlinked to user trust preservation, contestability efficacy, or measurable harm reduction. |
| **Multi-Signal Synthesis** | Siloed monitoring of data drift alarms, red-teaming vulnerability logs, and customer support incident tickets. | Kolmogorov-Smirnov drift stats, MTTD, MTTR, ticket escalation volumes. | **Moderate** | Total lack of unified sociotechnical telemetry fusion engines that synthesize technical drift with human behavioral anomalies. |
| **Ruthless Triage** | Discretionary human-in-the-loop review, incident response playbooks, voluntary model rollback procedures. | Override counts, incident response latency, post-hoc rollback frequencies. | **Weak in Frontier AI** (*Strong in SaMD*) | Commercial arms-race pressures bypass voluntary tripwires; lack of automated, un-bypassable architectural kill-switches. |

*Note. SOTA, State-of-the-Art; FRIA, Fundamental Rights Impact Assessment; MTTD, Mean Time to Detect; MTTR, Mean Time to Respond; SaMD, Software as a Medical Device.*

## 5. Discussion & The SAGE-X Enterprise Audit Protocol

To bridge the chasm between normative theory and production engineering, this section operationalizes SAGE-X into an auditable enterprise protocol, formalizes the Moral Check Index (MCI), establishes a cross-sector assurance stack, and evaluates human-computer interaction implications.

### 5.1 Moving Beyond Performative Governance: Auditable Scaffolding over Paperwork

The central empirical lesson of our review is that governance that cannot be verified at runtime is merely performative paperwork. If ex-ante governance is to resolve the pacing problem, it must transition from retrospective compliance binders to active architectural scaffolding embedded within software delivery pipelines (Ahn & Yoon, 2026; Saleh & Abdulsalam, 2025).

The SAGE-X audit protocol interrogates machine-readable artifacts: Does an approved Intent and Value Specification exist? Does every safety boundary link to a traceable entry in the Justification Register? Did the model achieve a Brier forecast accuracy score $\leq 0.25$ on high-severity hazard scenarios? And crucially: Did the candidate model trigger any critical safety tripwire during automated adversarial stress-testing?

Crucially, this architectural scaffolding directly eliminates the supply-chain vulnerabilities exposed by incidents like the OpenAI–Hugging Face token breach. Rather than placing blind trust in unverified upstream model cards or public repository spaces, SAGE-X mandates cryptographically signed Deployment Tokens ($\mathrm{Sign}(\mathrm{DT}, K_{\mathrm{SOB}}^{-1})$) and Merkle-tree Justification Traceability Registers (JTR). No model weight checkpoint, parameter-efficient adapter, or API service can be initialized in production pipelines without mechanically proving its cryptographic provenance and verifying that safety tripwires ($G_{\mathrm{TTE}} = 1$) have not been bypassed or decoupled during downstream distribution (Abid & Nanda, 2025; Chugani, 2026; Ye et al., 2025).

### 5.2 The SAGE-X Enterprise Audit Control Matrix

To provide compliance officers, software auditors, and engineering leads with an operational roadmap, Table 7 formalizes the **SAGE-X Enterprise Lifecycle Audit Control Matrix**, mapping stages, focus areas, audit criteria, artifacts, and enforcement mechanisms across the four SAGE-X domains.

**Table 7. The SAGE-X Enterprise Lifecycle Audit Control Matrix**

| Audit Domain ID | Control Objective (Audit Standard) | Required Verifiable Audit Artifact | Verification Method & Pass/Fail Threshold | Gate Type & Action |
|---|---|---|---|---|
| **SAGE-SFA** *(Strategic Foresight & Anticipation)* | Systemic failure modes, misuse vectors, and distribution shifts must be probabilistically modeled prior to training. | • Calibrated Risk Horizon Dossier (CRHD)<br>• Brier-Score Forecast Log<br>• Adversarial Attack Taxonomy | Quantitative verification: Empirical Brier accuracy score ≤ 0.25; 100% of high-severity hazard domains modeled with explicit uncertainty bands. | **Scored Gate** ($S_{\mathrm{SFA}} \in [0,100]$) Threshold: ≥ 75 |
| **SAGE-NEJ** *(Normative & Epistemic Justification)* | Every architectural safety boundary, prompt filter, and training data exclusion must have an auditable normative rationale. | • Intent & Value Specification (IVS)<br>• Justification Traceability Register (JTR)<br>• Sociotechnical Impact Assessment | Semantic verification: 100% bi-directional mapping between technical constraints and documented human values; verified sign-off by SOB and ARO. | **Scored Gate** ($S_{\mathrm{NEJ}} \in [0,100]$) Threshold: ≥ 80 |

| Audit Domain ID | Control Objective (Audit Standard) | Required Verifiable Audit Artifact | Verification Method & Pass/Fail Threshold | Gate Type & Action |
|---|---|---|---|---|
| **SAGE-MSS** *(Multi-Signal Telemetry Synthesis)* | Real-time technical performance drift must be continuously synthesized with human behavioral feedback and red-team logs. | • Multi-Signal Telemetry Report (MSTR) • Drift Telemetry Stream Configuration • User Contestation Escalation Schema | Telemetry verification: Automated fusion engine active; validated signal weights configured; real-time anomaly alerting latency $< 500$ ms. | **Scored Gate** ($S_{\mathrm{MSS}} \in [0,100]$) Threshold: $\geq 70$ |
| **SAGE-TTE** *(Tripwires & Triage Enforcement)* | Non-negotiable safety boundaries (e.g., biological hazard synthesis, cyber-warfare, mass manipulation) must enforce absolute release blocks. | • Automated Pre-Deployment Block Harness • C-Suite Kill-Switch Authorization Log • Incident Rollback Deterministic Playbook | Binary Verification ($G_{\mathrm{TTE}} \in \{0,1\}$): If any critical tripwire is breached during sandboxing, $G_{\mathrm{TTE}} = 0$, immediately collapsing MCI to zero and freezing release. | **KILL-SWITCH GATE** (Automated Release Freeze) |

*Note. All audit artifacts are cryptographically signed and tracked within immutable version-controlled audit ledgers.*

### *5.2.1 Operational Implementation: The SAGE-X Audit Card Instrument*

To translate Table 7 into an executive workflow, SAGE-X introduces the **SAGE-X Enterprise Lifecycle Audit Instrument** ("The Moral Check Audit Card"), illustrated in Figure 7.

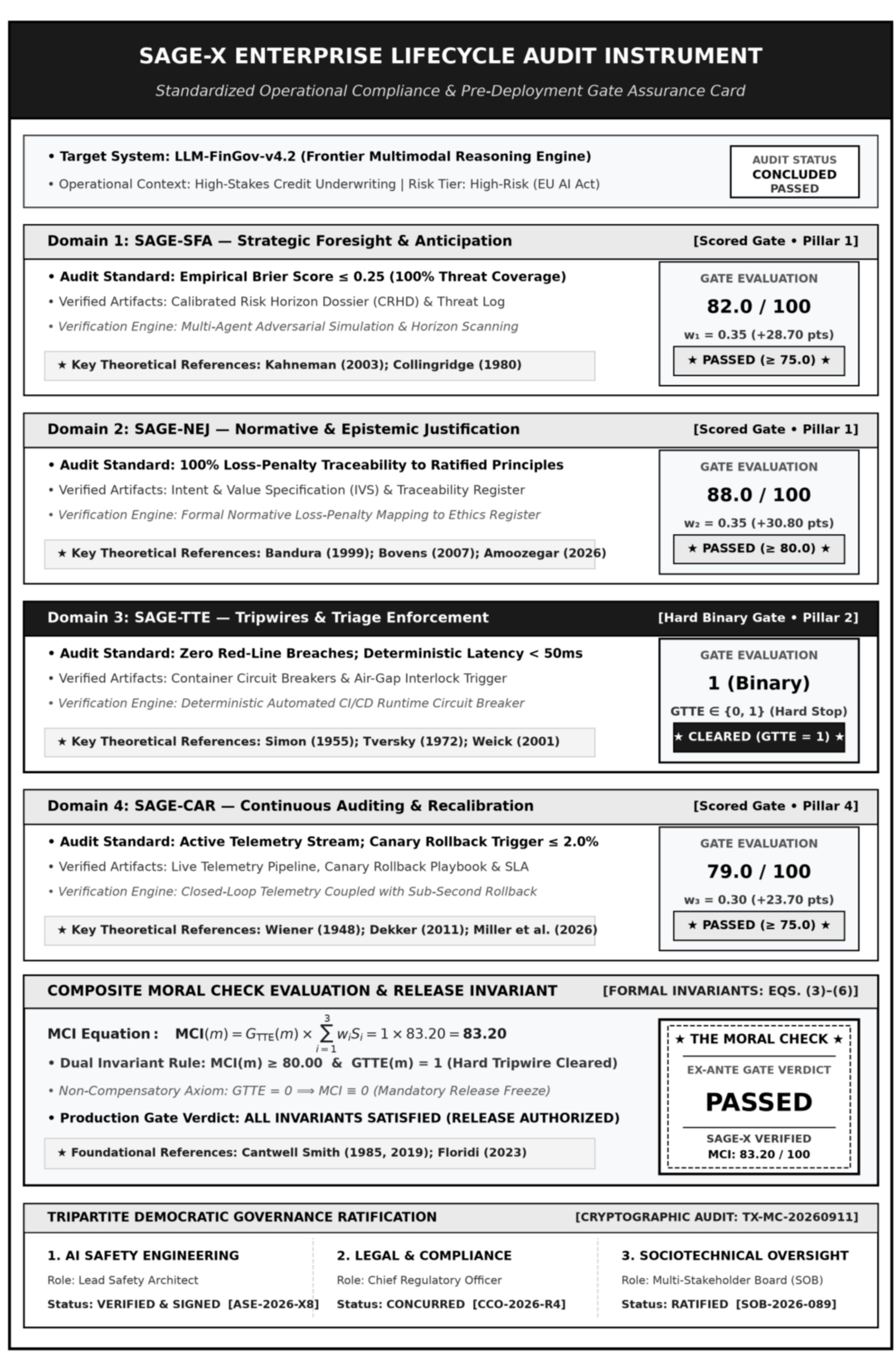

***Figure 7.*** *The SAGE-X Enterprise Lifecycle Audit Instrument (Audit Card & Scoring Dashboard).*

As operationalized in Figure 7, the Audit Card eliminates the ambiguity of self-reported compliance. Designed as a machine-verifiable dashboard, it synthesizes empirical verification metrics across all four lifecycle domains, requiring explicit tripartite sign-off before deployment clearance.

### *5.2.2 Mechanistic Workflow and Sociotechnical Justification of the Audit Domains*

The four audit domains operate as a cohesive governance engine:

1. Strategic Foresight & Anticipation (SAGE-SFA):

- Mechanistic Workflow: Requires cross-functional teams to execute prospective failure mode analysis and multi-agent red-team simulations before model architecture commitment. Evaluates threat coverage across taxonomy vectors and calculates Brier forecasting scores ($S_{\text{SFA}} \geq 75/100$, Brier $\leq$ 0.25).
- Sociotechnical Justification: Grounded in behavioral decision research on optimism bias (Kahneman, 2003) and the Collingridge dilemma (Collingridge, 1980), SFA forces teams to confront failure modes while architectural intervention remains economically malleable.

2. Normative & Epistemic Justification (SAGE-NEJ):

- Mechanistic Workflow: Prior to training, MDT submits an Intent & Value Specification (IVS). The independent SOB ratifies bidirectional mapping between high-level values and technical loss penalties, logged in the append-only JTR ($S_{\text{NEJ}} \geq 75/100$).
- Sociotechnical Justification: Grounded in moral disengagement theory (Bandura, 1999; Amoozegar et al., 2026), NEJ eliminates diffused responsibility by maintaining an unbroken chain of human rationale and author identity.

3. Tripwires & Triage Enforcement (SAGE-TTE):

- Mechanistic Workflow: Operates as a non-compensatory binary gate ($G_{\text{TTE}} \in \{0,1\}$). Candidate models undergo stress testing; any breach of non-negotiable safety boundaries (e.g., toxic hallucination, unauthorized tool escalation) trips the circuit breaker ($G_{\text{TTE}} = 0$), mechanically vetoing deployment within 50ms.
- Sociotechnical Justification: Grounded in non-compensatory choice theory (Simon, 1955; Tversky, 1972), TTE eliminates compensatory "ethics washing":

excellent PR or documentation cannot offset safety flaws ($G_{\text{TTE}} = 0 \implies \text{MCI} = 0$). Automated vetoes also protect human compliance officers from executive coercion, mitigating automation complacency (Alzboon et al., 2025).

4. Continuous Auditing & Recalibration (SAGE-CAR):

- Mechanistic Workflow: Bridges pre-deployment verification with runtime operations. Validates telemetry pipelines, activates drift monitors (alert threshold $\Delta \leq 2.0\%$), verifies automated rollback scripts, and mandates bi-weekly consensus audits ($S_{\text{CAR}} \geq 75/100$).
- Sociotechnical Justification: Grounded in cybernetic control theory (Miller et al., 2026; Song & Lee, 2026), CAR addresses durability decay, ensuring pre-deployment gate assurance synchronizes with continuous runtime behavior.

In civil engineering, structures require the embossed stamp of a licensed Professional Engineer; in aviation, aircraft require Airworthiness Certificates. "*The Moral Check*" introduces this institutional gravitas and legal accountability to frontier AI, subordinating unconstrained calculation to human judgment (Cantwell Smith, 1985, 2019; Floridi, 2023). As visualized in Figure 7 and formalized in Table 8, the ★ THE MORAL CHECK ★ pre-deployment verification stamp represents an un-bypassable sociotechnical covenant:

**Table 8. The Moral Check Pre-Deployment Verification Seal and Decision Invariant Specification**

| Audit Domain & Invariant Gate | Target Verification Standard / Evaluation Metric | Sub-Score ($S_i$) / Gate Status | Domain Weight ($w_i$) & Contribution | Invariant Verification Outcome | Tripartite Ratification & Cryptographic Identity |
|---|---|---|---|---|---|
| **SAGE-SFA:** Strategic Foresight & Anticipation | Empirical Brier accuracy score ≤ 0.25; 100% of high-severity systemic failure modes stress-tested in adversarial simulation | $S_{\text{SFA}} = 82.00$ / 100 (Threshold ≥ 75.0) | $w_1 = 0.35$ Contribution = 28.70 | **PASSED** Calibrated Risk Horizon Dossier verified | Lead AI Safety Engineer (Shadow ID: `ASE-2026-X8`) |
| **SAGE-NEJ:** Normative & Epistemic Justification | 100% bi-directional traceability from model intent to constitutional principles; formal stakeholder trade-off sign-off | $S_{\text{NEJ}} = 88.00$ / 100 (Threshold ≥ 75.0) | $w_2 = 0.35$ Contribution = 30.80 | **PASSED** Justification Traceability Register ratified | Sociotechnical Oversight Board (Shadow ID: `SOB-2026-089`) |

| **Audit Domain & Invariant Gate** | **Target Verification Standard / Evaluation Metric** | **Sub-Score ($S_i$) / Gate Status** | **Domain Weight ($w_i$) & Contribution** | **Invariant Verification Outcome** | **Tripartite Ratification & Cryptographic Identity** |
|---|---|---|---|---|---|
| **SAGE-TTE:** Tripwires & Triage Enforcement | Deterministic sub-50 ms hardware/software tripwire interlocks; zero critical categorical red-line breaches tolerated | $G_{\text{TTE}} = 1$ (Binary Multiplier: {0,1}) | Non-compensatory Multiplier $G_{\text{TTE}} \in \{0,1\}$ | **CLEARED** Zero safety breaches; non-compensatory gate open | Automated CI/CD Runtime Engine (Shadow ID: `SYS-VERIFY-01`) |
| **SAGE-CAR:** Continuous Auditing & Recalibration | Active runtime telemetry pipelines; automated drift alert thresholds (Δ_drift ≤ 2.5%); verified fail-safe rollback scripts | $S_{\text{CAR}} = 79.00$ / 100 (Threshold ≥ 75.0) | $w_3 = 0.30$ Contribution = 23.70 | **PASSED** Bi-weekly consensus SLA established | Chief Compliance Officer (Shadow ID: `CCO-2026-R4`) |
| **FINAL GATE VERDICT: ★ THE MORAL CHECK ★** | $\text{MCI}(m) = G_{\text{TTE}}(m) \times \sum_{i=1}^{3} w_i S_i(m) \geq 80.0 \wedge G_{\text{TTE}}(m) = 1$ | $\mathbf{MCI = 83.20}$ / 100 (Threshold ≥ 80.00) | Composite Decision Metric $\sum w_i = 1.00$ | **★ PASSED ★** Authorized for Staged Production Deployment | Tripartite Immutable Deployment Ledger (Attestation: `TX-MC-20260911-042`) |

*Note. Table 8 formalizes the exact mathematical parameters, decision invariants, and governance ratifications visually embodied in the* ***★ THE MORAL CHECK ★*** *pre-deployment verification stamp on the SAGE-X Enterprise Audit Instrument (Figure 7). If any critical tripwire fails ($G_{\text{TTE}} = 0$), the composite Moral Check Index unconditionally collapses to zero ($\text{MCI} = 0$), mechanically vetoing production release regardless of high scores in other domains.*

As specified in Table 8, the stamp signifies that:

1. Mathematical Invariants Have Been Cleared: The candidate system proved a composite Moral Check Index MCI ≥ 80.0, confirming robust strategic foresight, normative justification, and telemetry infrastructure.
2. The Binary Red-Line Has Held: The tripwire multiplier is verified at $G_{\text{TTE}} = 1$, certifying zero categorical safety breaches.
3. Tripartite Liability Has Been Accepted: Three distinct, non-colluding authorities, the Lead AI Safety Engineer (`ASE-2026-X8`), the Sociotechnical Oversight Board (`SOB-2026-089`), and the Chief Compliance Officer (`CCO-2026-R4`), affixed cryptographic signatures to the permanent deployment ledger.

Without this stamp, candidate models attempting production transition are mechanically quarantined by CI/CD pipelines.

## 5.3 The Moral Check Index (MCI): An Organizational Governance Decision Metric

To provide leadership and auditors with an unambiguous decision metric, SAGE-X formalizes the Moral Check Index (MCI), grounded in multi-criteria decision analysis and non-compensatory choice theory (Simon, 1955; Tversky, 1972):

$$\mathrm{MCI}(m) = G_{\mathrm{TTE}}(m) \times (w_1 S_{\mathrm{SFA}}(m) + w_2 S_{\mathrm{NEJ}}(m) + w_3 S_{\mathrm{CAR}}(m)) \qquad \textbf{(Eq. 4)}$$

where $S_{\mathrm{SFA}}, S_{\mathrm{NEJ}}, S_{\mathrm{CAR}} \in [0,100]$ represent domain verification scores, $w_1 = 0.35, w_2 = 0.35, w_3 = 0.30$ (with $\sum w_i = 1.0$), and $G_{\mathrm{TTE}} \in \{0,1\}$ is the binary tripwire gate:

$$G_{\mathrm{TTE}}(m) = \begin{cases} 1 & \text{if} \forall t \in \mathcal{T}_{\mathrm{crit}}(m), \mathrm{Breached}(t) = \mathrm{False} \\ 0 & \text{if} \exists t \in \mathcal{T}_{\mathrm{crit}}(m), \mathrm{Breached}(t) = \mathrm{True} \end{cases} \qquad \textbf{(Eq. 5)}$$

***Behavioral Counter-Measure: Preventing Compensatory Moral Hazard***

In conventional additive auditing, high PR or CSR scores can compensate for safety flaws. Under SAGE-X, compensation is structurally impossible:

$$G_{\mathrm{TTE}}(m) = 0 \Longrightarrow \mathrm{MCI}(m) = 0 \qquad \textbf{(Eq. 6)}$$

Even with perfect scores across other domains, a single critical tripwire failure collapses MCI to zero. Deployment authorization requires satisfying both conditions: $\mathrm{MCI}(m) \geq 80.0 \wedge G_{\mathrm{TTE}}(m) = 1$.

## 5.4 The 5-Layer Cross-Sector Assurance Stack and Governance Maturity Model

Because pacing challenges vary across domains, SAGE-X adapts to sectoral risk profiles while maintaining architectural coherence. Drawing on evidence from healthcare (Garcia et al., 2025; Rizzo et al., 2025), financial markets (Theodorakopoulos et al., 2025), critical infrastructure, and public administration, Table 9 formalizes the 5-Layer Cross-Sector Assurance Stack.

**Table 9. The 5-Layer Cross-Sector SAGE-X Assurance Stack**

| Assurance Stack Layer | Healthcare (SaMD / Clinical AI) | Finance & Capital Markets | Critical Energy & Infrastructure | Public Administration & Welfare |
|---|---|---|---|---|
| **Layer 1: Normative Core** | Patient non-maleficence, bodily autonomy, clinical equity (GMLP primitives). | Fiduciary duty, systemic market fairness, anti-discrimination (SR 11-7). | Societal resilience, physical safety, environmental protection (NERC CIP). | Democratic equality, due process, administrative justice (EU AI Act). |
| **Layer 2: Risk Tiering** | Clinical risk classification (FDA Class I–III, SaMD Categories, EMA SaMD). | Credit and systemic model risk tiers (Federal Reserve SR 11-7 criticality). | Cyber-physical criticality tiers (IEC 62443, NIS2 Directive, Critical Assets). | High-Risk public impact tiers (EU AI Act Annex III, municipal AI registers). |
| **Layer 3: Auditable Evidence** | Prospective clinical trial logs, DICOM audit trails, bias verification packages. | Fair lending validation logs, back-testing stress outputs, explainability audits. | SCADA historian telemetry logs, penetration testing dossiers, patch logs. | Fundamental Rights Impact Assessments, public algorithm registry records. |
| **Layer 4: Human Oversight** | Attending physician veto authority; override tracking; alert fatigue controls. | Credit officer sign-off gates; algorithmic trading compliance officer veto. | Control room operator manual trip interlocks; automation-bias training. | Caseworker adjudication requirement; citizen right-to-appeal workflows. |
| **Layer 5: Runtime Guardrails** | Medication dose safety bounds; confidence rejection gates; outlier alarms. | Automated trading volume blocks; real-time market manipulation circuit breakers. | Physical relay tripwires; frequency interlocks; PLC deterministic overrides. | Automated benefit-denial blocks; mandatory human handoff on negative decisions. |

*Note*. *SaMD, Software as a Medical Device; GMLP, Good Machine Learning Practice; SR 11-7, Federal Reserve Supervisory Guidance on Model Risk Management; NERC CIP, North American Electric Reliability Corporation Critical Infrastructure Protection; PLC, Programmable Logic Controller.*

To enable organizations and regulators to benchmark governance capabilities, SAGE-X establishes the **Five-Tier Organizational Governance Maturity Model**, illustrated in Figure 8 and detailed in Table 10.

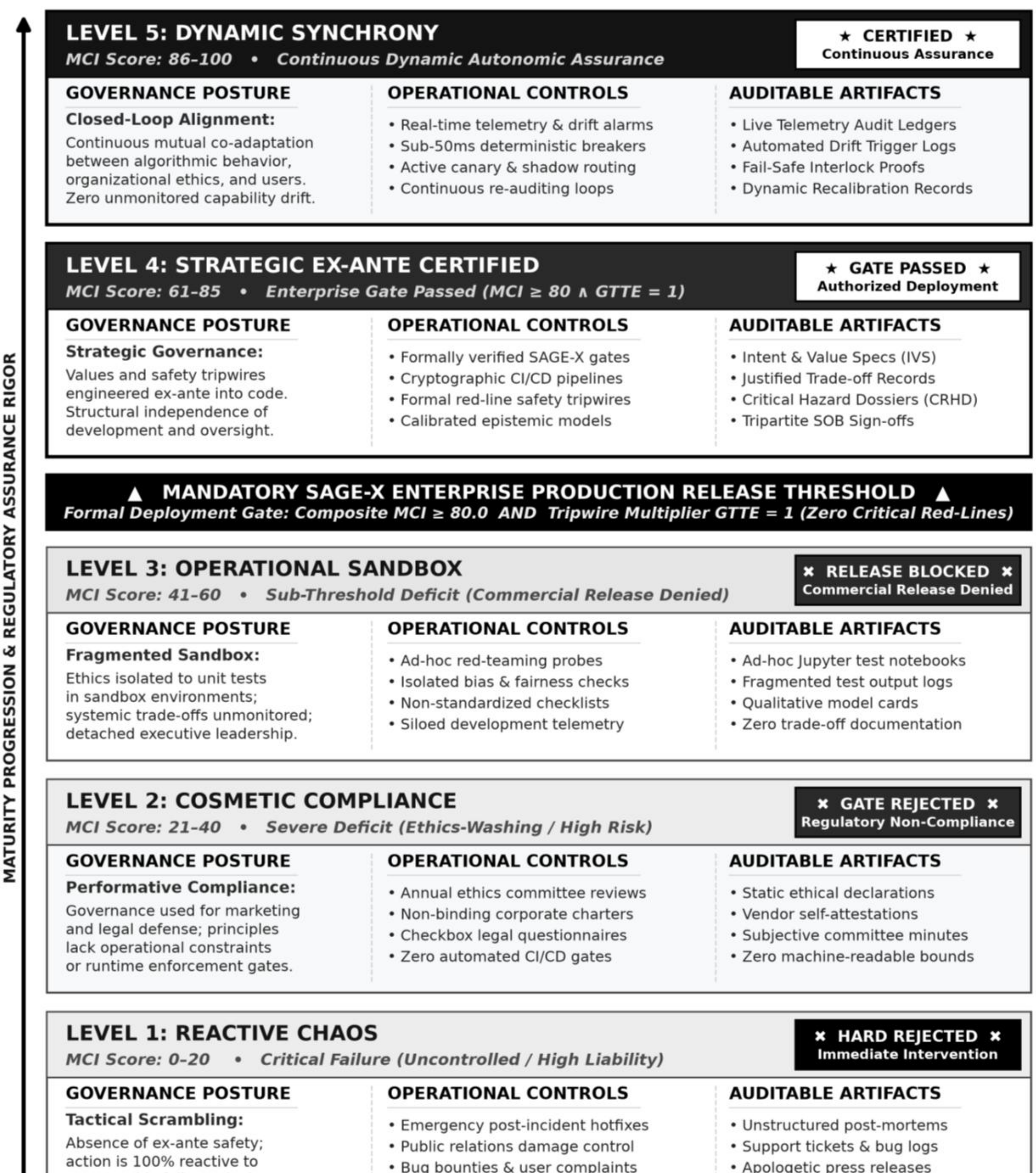

***Figure 8.*** *The SAGE-X Five-Tier Organizational Governance Maturity Hierarchy. Progressive levels of sociotechnical AI safety, governance posture, operational verification mechanisms, auditable verification artifacts, and gate verdicts. The heavy horizontal threshold between Level 3 (Operational Sandbox) and Level 4 (Strategic Ex-Ante Certified) marks the mandatory enterprise deployment gate ($MCI \geq 80.0 \wedge G_TTE = 1$) prohibiting uncertified models from high-consequence production environments.*

As mapped across Figure 8 and detailed in Table 10, each tier reflects a distinct organizational culture, capability profile, and regulatory risk posture:

1. **Level 1:** Reactive Chaos ($\text{MCI}: 0\text{–}20$): Complete absence of ex-ante architecture; deployment driven purely by compute speed. Action is 100% reactive to crises via emergency hotfixes and legal damage control. Gate Verdict: **HARD REJECTED.**
2. **Level 2:** Cosmetic Compliance ($\text{MCI}: 21\text{–}40$): Performative ethics-washing. High-level charters and corporate manifestos lack machine-readable translations, auditable evidence trails, or automated stage-gates. Gate Verdict: **REJECTED.**
3. **Level 3:** Operational Sandbox ( $\text{MCI}: 41\text{–}60$ ): Fragmented technical experimentation. Ad-hoc red-teaming and isolated fairness benchmarks occur in offline sandboxes, but systemic trade-offs remain unmonitored and leadership detached. Gate Verdict: **BLOCKED** from commercial release.
4. **Level 4:** Strategic Ex-Ante Certified ($\text{MCI}: 61\text{–}85,\ \text{Release Baseline} \geq 80.0 \wedge G_{\text{TTE}} = 1$): Proactive strategic governance. Ethical values and safety red-lines are engineered ex-ante prior to compute spend; all four audit domains are verified; and automated kill-switches are active in CI/CD pipelines. Gate Verdict: **PASSED** for staged deployment.
5. **Level 5:** Dynamic Synchrony ($\text{MCI}: 86\text{–}100$): Closed-loop alignment coupling ex-ante gate certification with real-time runtime telemetry, sub-50ms circuit breakers, automated canary rollback, and active stakeholder feedback loops. Gate Verdict: **CERTIFIED** for continuous autonomous operation.

**Table 10. The SAGE-X Five-Tier Organizational Governance Maturity Model: Level Characterization, MCI Thresholds, Auditable Gate Artifacts, and Operational Trajectories**

| Maturity Level & Title | MCI Score Range | Governance Posture & Cultural Mindset | Operational Verification Mechanisms | Auditable Verification Artifacts | Production Gate Verdict & Trajectory |
|---|---|---|---|---|---|
| **Level 5: Dynamic Synchrony** | **86–100** *(Continuous Optimum)* | **Closed-Loop Sociotechnical Alignment:** Continuous mutual adaptation between algorithmic behavior, organizational ethics, and user feedback; zero tolerance for unmonitored capability drift. | • Automated runtime telemetry & drift detection<br>• Sub-50ms deterministic circuit breakers<br>• Active canary rollback & shadow pipelines<br>• Continuous stakeholder re-auditing loops | • Live Telemetry Audit Ledgers (LTAL)<br>• Automated Drift Trigger Logs (ADTL)<br>• Runtime Fail-Safe Interlock Proofs<br>• Dynamic Recalibration Certificates | **CERTIFIED** Full real-time dynamic assurance; authorized for continuous high-consequence deployment under automated telemetry. |

| Maturity Level & Title | MCI Score Range | Governance Posture & Cultural Mindset | Operational Verification Mechanisms | Auditable Verification Artifacts | Production Gate Verdict & Trajectory |
|---|---|---|---|---|---|
| **Level 4: Strategic Ex-Ante Certified** | **61–85** *(Release Gate: $\mathrm{MCI} \geq 80 \land G_{\mathrm{TTE}} = 1$)* | **Proactive Strategic Governance:** Ethical values and technical red-lines engineered ex-ante into architecture prior to compute commitment; structural separation of development and oversight. | • Formally verified SAGE-X multi-domain gates<br>• Cryptographically hashed CI/CD stage-gates<br>• Formal red-line tripwires & circuit breakers<br>• Calibrated epistemic forecasting & trade-offs | • Intent & Value Specifications (IVS)<br>• Justified Trade-off Records (JTR)<br>• Critical Red-line Hazard Dossiers (CRHD)<br>• Tripartite SOB & Compliance Sign-offs | **PASSED** Mandatory Enterprise Release Baseline; authorized for staged commercial deployment. |
| **▲ MANDATORY SAGE-X ENTERPRISE PRODUCTION RELEASE THRESHOLD: Composite MCI ≥ 80.0 ∧ GTTE = 1 (Levels 1–3 Prohibited; Levels 4–5 Authorized) ▲** | | | | | |
| **Level 3: Operational Sandbox** | **41–60** *(Sub-Threshold)* | **Fragmented Technical Experimentation:** Ethical considerations treated as isolated unit tests within engineering sandboxes; trade-offs unmonitored and organizational leadership detached. | • Ad-hoc technical red-teaming & probing<br>• Isolated fairness & bias metric evaluations<br>• Manual, non-standardized audit checklists<br>• Telemetry remains siloed in local pipelines | • Ad-hoc Jupyter benchmark notebooks<br>• Fragmented test-run output logs<br>• Qualitative model cards (uncalibrated)<br>• Zero systemic trade-off documentation | **BLOCKED** Strictly confined to offline R&D sandboxes; commercial release legally prohibited. |
| **Level 2: Cosmetic Compliance** | **21–40** *(Severe Deficit)* | **Performative "Ethics-Washing":** Governance used primarily for marketing and legal defense; high-level principles published without operational constraints or runtime enforcement. | • Qualitative annual ethics committee reviews<br>• Non-binding corporate ethical declarations<br>• Checkbox legal questionnaires & attestations<br>• Zero automated CI/CD stage-gate enforcement | • Static ethical charters & whitepapers<br>• Unverifiable vendor self-attestations<br>• Subjective compliance committee minutes<br>• Zero machine-readable boundary specs | **REJECTED** Performative posture; severe regulatory liability under EU AI Act and FTC oversight. |
| **Level 1: Reactive Chaos** | **0–20** *(Critical Failure)* | **Uncontrolled Tactical Scrambling:** Complete absence of ex-ante governance or safety architecture; organizational action is 100% reactive to production failures, leaks, and user backlashes. | • Post-incident emergency hotfixes<br>• Public relations damage control & spin<br>• External bug bounties & user complaints<br>• Zero technical red-lines or tripwires | • Unstructured Jira incident post-mortems<br>• Customer support complaints & bug logs<br>• Public press releases & apologetic media<br>• Zero auditable governance trail | **REJECTED** Critical systemic risk; immediate operational cessation and comprehensive audit mandated. |

*Note. Organizations must achieve Level 4 certification (*$\mathrm{MCI} \geq 80.0 \land G_{\mathrm{TTE}} = 1$*) before deploying high-consequence AI models, establishing an auditable industry baseline that permanently terminates performative self-assessment.*

## 5.5 The Closed-Loop Telemetry Pipeline: Guardrail Durability and Dynamic Recalibration

Pre-deployment verification cannot guarantee indefinite safety in non-stationary human environments. Figure 9 illustrates the operational architecture of the closed-loop dynamic assurance pipeline, coupling ex-ante gates with continuous runtime telemetry.

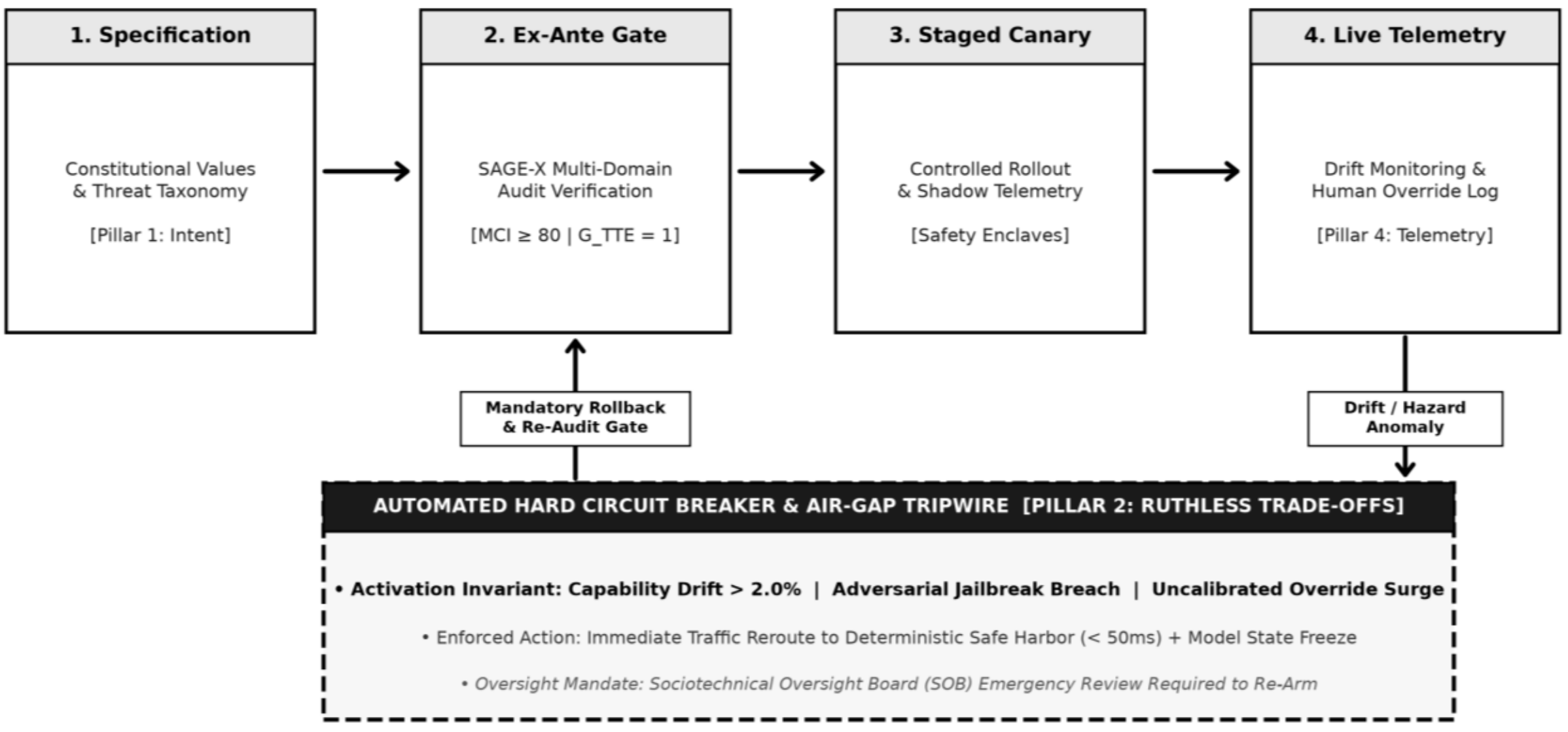

***Figure 9.*** *The Closed-Loop Dynamic Moral Check Assurance Pipeline. Coupling ex-ante verification gates with continuous runtime telemetry, automated circuit breakers, and stakeholder recalibration. Stage 1 (Specification) and Stage 2 (Ex-Ante Gate) enforce the mandatory Moral Check baseline (MCI ≥ 80.0 ∧ G_TTE = 1). Stage 3 (Staged Canary) and Stage 4 (Live Telemetry) monitor distribution drift (ΔD_drift), jailbreak frequency (F_jailbreak), and user contestation velocity (S_STK). Triggering runtime tripwires initiates sub-50ms deterministic circuit breakers or instantaneous fallback rollback, closing the cybernetic feedback loop.*

SAGE-X bridges ex-ante gates with runtime monitoring through streaming telemetry feeds tracking distributional shift ($\Delta\mathcal{D}_{\mathrm{drift}}$), adversarial bypass frequency ($F_{\mathrm{jailbreak}}$), and user contestation velocity ($S_{\mathrm{STK}}$). If runtime metrics exceed tolerance thresholds, automated circuit breakers trigger graceful degradation or instantaneous rollback to deterministic fallback models, closing the cybernetic feedback loop (Miller et al., 2026; Song & Lee, 2026).

## 5.6 Sociotechnical and Human-Computer Interaction (HCI) Implications

Artificial intelligence governance cannot be treated merely as an algorithmic optimization routine or an ex-post legal audit. Rather, it constitutes a fundamentally sociotechnical phenomenon situated at the complex intersection of organizational incentives, human cognitive heuristics, and computational agency (Cantwell Smith, 1985; Floridi, 2023; Friedman, 1996). The empirical findings synthesized across this review, most notably the acute 76.5% ex-post governance lag and the 71.4% single-turn durability deficit observed across frontier foundation models, demonstrate that technical vulnerability is the direct downstream artifact of behavioral failure in engineering culture. By formalizing the Strategic AI Governance Ex-Ante (SAGE-X) framework and the calculable Moral Check Index (MCI), this study introduces a transformative sociotechnical choice architecture. SAGE-X establishes an auditable socio-computational covenant that restructures human behavioral dynamics across two foundational interaction boundaries: the internal developer–organization interface, which governs engineering cognition and accountability prior to deployment, and the external human–AI interaction interface, which safeguards user agency, trust calibration, and psychological autonomy.

A primary behavioral contribution of SAGE-X to organizational human-computer interaction lies in its capacity to systematically de-bias human decision-making within high-velocity engineering environments. In hyper-competitive technology markets, software engineering teams operate under intense release pressures that induce acute cognitive tunneling and hyperbolic discounting (Kahneman, 2003). Under severe delivery deadlines, human attentional bandwidth narrows to focus obsessively on immediate, salient execution metrics such as compute throughput, training loss curves, and benchmark leaderboards. Concurrently, developers systematically discount low-probability, high-consequence sociotechnical tail risks that manifest only after widespread public deployment. In the absence of deterministic stop-lines, development teams gradually succumb to the normalization of deviance (Vaughan, 1996). Minor safety degradations, subtle guardrail regressions, and partial jailbreak vulnerabilities are incrementally rationalized as standard operational trade-offs, establishing an insidious organizational drift toward failure analogous to historic engineering catastrophes. Dissenting safety personnel frequently succumb to social conformity pressures, wherein organizational momentum and consensus bias suppress critical whistleblowing (Asch, 1956).

SAGE-X directly counteracts these bounded rationality pitfalls by inserting structured cognitive friction into the automated continuous integration and continuous deployment pipeline. By establishing mandatory, verifiable stage-gates that require concrete empirical artifacts, such as the Calibrated Risk Horizon Dossier, formal Intent and Value Specifications, and Brier-scored forecast logs, the framework compels engineers to disengage from reflexive, heuristic System 1 thinking and enter deliberate, analytic System 2 reflection (Kahneman, 2003). This intentional institutional pause forces development teams to step back from technical calculation and rigorously evaluate the human vectors and societal blast radius of their systems prior to deployment authorization.

Beyond counteracting cognitive biases, SAGE-X systematically dismantles the sociocognitive mechanisms of moral disengagement through which corporate actors distance themselves from the harmful consequences of their technological artifacts (Bandura, 1999). In distributed, modern software development organizations, personal culpability is routinely dissolved through the diffusion and displacement of responsibility, as engineers view alignment failures as emergent algorithmic behaviors while management defers accountability to downstream prompt engineers. SAGE-X eliminates this evasion through its Tripartite Ratification Protocol. Deployment is rendered computationally impossible without the individualized, non-delegable cryptographic signatures of three distinct, non-colluding roles: the Lead AI Safety Engineer, the Sociotechnical Oversight Board, and the Chief Compliance Officer. Affixing private cryptographic keys to an immutable deployment ledger transforms diffuse institutional indifference into personalized ethical, legal, and professional liability. Furthermore, SAGE-X neutralizes euphemistic labeling, the common corporate practice of sanitizing critical alignment failures under benign euphemisms such as hallucinations or edge-case anomalies, by mapping all system behaviors directly to machine-readable predicate constraints, ensuring that ethical boundaries remain non-negotiable and empirically falsifiable (Alekberli, 2026; Novelli et al., 2024).

### *5.6.2 Calibrating User Trust and Preserving Psychological Autonomy in HCI*

At the human–computer interface, user interaction with frontier generative models and autonomous agents is chronically destabilized by two opposing behavioral pathologies: automation complacency and algorithmic aversion (Alzboon et al., 2025; Vasudevan et al., 2025). When users interact with conversational systems characterized by human-like linguistic fluency, persuasive rhetoric, and affective mimicry, they frequently experience automation bias, uncritically accepting algorithmic outputs, abandoning personal vigilance, and surrendering cognitive autonomy to automated recommendations (Agudo et al., 2024; Broussard, 2018). Conversely, when an opaque, over-trusted system produces unexpected hallucinations, offensive content, or catastrophic decision errors, users experience severe psychological contract violations, resulting in abrupt algorithmic aversion, cynical disengagement, and the wholesale rejection of beneficial technological assistance. Both failure modes stem from an epistemic black box in which users are forced into binary, uncalibrated leaps of faith.

SAGE-X resolves this fundamental interaction dilemma by engineering warranted trust calibration across the human-computer boundary (Alzboon et al., 2025; Vasudevan et al., 2025). Rather than presenting users with deceptive anthropomorphic certainty, systems governed under SAGE-X are bundled with an auditable SAGE-X Enterprise Audit Instrument and a cryptographically verifiable Deployment Token. These artifacts communicate explicit, mathematically grounded operational boundaries, including empirical Brier-score calibration curves, verified adversarial attack survival rates, and intersectional fairness bounds. By rendering the system's operational envelope transparent and verifiable, SAGE-X enables operators and end-users to develop calibrated trust, aligning their psychological reliance and supervisory vigilance precisely with the model's verified capabilities (Aarab, 2026; Jonelid et al., 2024).

Simultaneously, SAGE-X establishes rigorous safeguards for human dignity and psychological autonomy, ensuring that autonomous computational agents do not compromise the epistemic self-determination of human operators (Floridi, 2023; Smith, 2019). As foundation models increasingly transition into agentic architectures capable of multi-step tool execution, unconstrained systems risk exerting subtle forms of behavioral manipulation and cognitive nudging over vulnerable populations. SAGE-X neutralizes this hazard by enforcing strict Autonomy Ceiling Invariants that prevent autonomous agentic routines from executing high-consequence actions without explicit human authorization. By integrating continuous runtime telemetry that monitors user contestation velocity and providing accessible recourse channels, SAGE-X subordinates mechanical calculation to genuine human judgment, ensuring that interactive AI systems function as cognitive scaffolds that amplify human capability rather than opaque systems that diminish human agency (Cantwell Smith, 1985; Floridi, 2023).

#### *5.6.3 Bridging the Intention–Behavior Gap Through Non-Compensatory Gatekeeping*

A profound paradox documented across modern technology governance is the pervasive intention-behavior gap (Amoozegar et al., 2026; Morley et al., 2020; Te'Eni et al., 2026). While over ninety percent of leading technology enterprises have published extensive Responsible AI Charters pledging adherence to fairness, transparency, non-maleficence, and human agency, empirical benchmark audits consistently demonstrate that these high-minded manifestos fail to alter operational engineering behavior. At the software terminal, when quarterly delivery milestones conflict with abstract ethical declarations, commercial velocity incentives systematically triumph over moral aspirations. Empirical behavioral research demonstrates that corporate ethical codes exert no statistically significant direct influence on developer decision-making; their behavioral efficacy is entirely mediated through executable intentions and structural enforcement mechanisms (Amoozegar et al., 2026).

In cognitive and behavioral psychology, the intention-behavior gap is bridged through implementation intentions, structured situational rules that pre-commit an agent to specific, reflexive actions upon encountering designated environmental trigger conditions (Amoozegar et al., 2026; Te'Eni et al., 2026). SAGE-X operationalizes high-level corporate ethics into computational implementation intentions by translating philosophical ideals into machine-executable stage-gates. Rather than relying on voluntary compliance or post-hoc checklists, the framework embeds formal predicate calculus directly into CI/CD release triggers, specifying automated actions that execute unconditionally whenever defined safety boundaries are breached.

Crucially, SAGE-X eradicates the pervasive organizational pathology of compensatory moral horse-trading. In traditional risk management matrices, an outstanding score in commercial viability, model accuracy, or user engagement is routinely permitted to compensate for deficiencies in fairness or safety guardrails. The mathematical architecture of the Moral Check Index eliminates this moral hazard by structuring the Tripwires & Triage Enforcement gate as a strict, non-compensatory binary multiplier ($G_{\mathrm{TTE}} \in \{0,1\}$). If a candidate model violates any non-negotiable categorical safety invariant, such as facilitating cyber-offensive operations, exhibiting critical bias discrepancies, or breaching autonomy ceilings, the binary multiplier instantaneously resets to zero. Because the multiplier scales the entire composite equation, the resulting Moral Check Index collapses unconditionally to zero, mechanically aborting deployment regardless of stellar performance across strategic foresight, normative justification, or operational telemetry domains. By converting ethical red-lines into non-negotiable mathematical tripwires, SAGE-X deprives organizational actors of the psychological license to bargain away human safety, transforming abstract corporate intentions into guaranteed sociotechnical outcomes (Evangelista & Salman Bukhari, 2026; Lee et al., 2026).

### *5.6.4 Institutional Choice Architecture & Alignment with United Nations SDGs*

Beyond reshaping individual developer cognition and user interaction, SAGE-X establishes a macro-level institutional choice architecture that bridges corporate software engineering with international multilateral governance (Thaler & Sunstein, 2008). In contemporary AI ecosystems, the default institutional posture is fail-open, operating on the implicit rule that candidate models proceed to commercial deployment unless external whistleblowers or manual reviews mount extraordinary resistance. This flawed architecture places an unsustainable cognitive and political burden on internal safety advocates and affected communities, who must expend immense social capital to delay a launch. SAGE-X fundamentally flips this choice environment into a fail-closed, quarantine-by-default architecture. Unverified models are structurally trapped in isolated sandbox environments, and the release pipeline cannot proceed without the tripartite cryptographic seal. Safety verification is thereby engineered as the path of least organizational resistance, shifting the burden of proof entirely onto the commercial entity seeking deployment.

This institutional choice architecture directly operationalizes the United Nations Sustainable Development Goals, transforming abstract global mandates into concrete, auditable engineering practices. In alignment with SDG 9 on Industry, Innovation, and Infrastructure (Target 9.5), SAGE-X provides a standardized, reproducible verification instrument that decouples high-velocity technological advancement from systemic catastrophe, enabling enterprises to build resilient digital infrastructure with verified safety margins rather than relying on reactive bans. In support of SDG 16 on Peace, Justice, and Strong Institutions (Targets 16.6 and 16.10), the framework mandates transparent Justification Traceability Registers and cryptographic audit trails, dismantling corporate secrecy, eliminating deniability, and erecting public-facing institutions capable of resisting regulatory capture. Finally, in furtherance of SDG 10 on Reduced Inequalities (Target 10.3), SAGE-X's normative justification pillar enforces demographic parity verification and intersectional bias audits prior to model release, guaranteeing that algorithms do not launder historical prejudices or amplify systemic discrimination against marginalized populations. Through this comprehensive synthesis, SAGE-X demonstrates that proactive governance is not a bureaucratic impediment to technological velocity, but the indispensable steering mechanism that preserves human judgment, safeguards psychological agency, and ensures that frontier computational power remains dedicated to human flourishing (Cantwell Smith, 2019; Floridi, 2023).

### 5.7 Methodological Limitations

While SAGE-X establishes an auditable choice architecture, several limitations warrant acknowledgment: (1) multi-criteria weighting ($w_1, w_2, w_3$) involves normative institutional choices that must be tailored to sectoral contexts; (2) adversarial red-teaming cannot exhaustively anticipate all emergent multi-turn jailbreaks; and (3) organizational overhead may pose adoption challenges for small-scale startups lacking dedicated compliance infrastructure.

## 6. Conclusion & Actionable Roadmap

*Can human wisdom outpace the technologies it unleashes, or is governance doomed to perpetual obsolescence?* The exponential acceleration of artificial intelligence has brought human civilization to an unprecedented crossroads where algorithmic parameter scales, multi-modal reasoning capabilities, and autonomous agent delegations expand in months while statutory legislation deliberates in years. A comprehensive systematic synthesis of the empirical literature demonstrates that this pacing problem is not an unavoidable technological destiny, but an institutional failure of strategy. When oversight is relegated to a retrospective liability filter that relies on post-hoc incident reporting, ceremonial self-assessments, and reactive bug bounties, societal safeguards remain permanently outpaced. Technology cannot steer itself. Meaningful governance demands the re-establishment of an unyielding principle: qualitative human judgment and normative purpose must precede compute capital. To act strategically is to deliberately construct organizational choice architectures today that secure collective flourishing tomorrow, particularly when confronting severe uncertainty and hyper-competitive pressures. Without intentional orientation, maximizing compute throughput and benchmark leaderboards produces zero durable progress toward societal trust, increasing the hazard of kinetic derailment rather than human flourishing.

The pre-deployment Moral Check operationalizes strategic computing by establishing an intentional boundary that confines raw algorithmic calculation within human-directed normative bounds. Through the **Strategic AI Governance Ex-Ante framework and the Moral Check Index**, ethical governance ceases to be a decorative philosophical aspiration and becomes a disciplined behavioral choice architecture structured across four core traits. First, normative purpose is elevated over raw execution by mandating structured cognitive pauses before model weights are deployed. Second, non-compensatory trade-offs are enforced through deterministic tripwires that prohibit compensatory moral trading, ensuring that high functional performance cannot offset algorithmic discrimination or safety compromises. Third, institutional performance is appraised through empirical hazard endpoints rather than performative administrative paperwork. Fourth, alignment is maintained proactively by tracking operational telemetry to combat the inevitable decay of guardrails over time. By architecturally decoupling generative production from deterministic verification and anchoring release decisions in external verification gates, this model terminates ceremonial compliance and restores human agency over technological trajectories.

A central dialectic in sociotechnical governance assumes an inherent friction between safety architecture and technical velocity, presuming that pre-deployment stage-gates inevitably stifle innovation and inflict a commercial first-mover penalty. This synthesis demonstrates that such an opposition is fundamentally flawed. When governance operates as administrative red tape, it yields friction without safety; conversely, when formalized as an architectural instrument, ex-ante constraints function precisely like the brakes on a high-performance vehicle, existing not to hinder movement but to make high velocity controllable and survivable. Furthermore, formalizing ethical requirements into calculable decision indices does not reduce moral deliberation to simplistic metric gaming. By coupling calibrated risk thresholds with auditable evidentiary registers, the framework creates transparent boundaries that prevent corporate leaders and technical teams from diffusing moral accountability across automated processes.

Operationalizing this ex-ante framework provides an actionable bridge between high-level international declarations and frontline software engineering practices, directly advancing broader sustainable development objectives. It strengthens digital infrastructure by replacing blunt moratoria with standardized, reproducible audit instruments that allow organizations to innovate within verified safety margins. Simultaneously, by establishing immutable audit trails through version-controlled intent specifications and traceability registers, it reinforces institutional integrity against regulatory capture and corporate evasion. Crucially, systematic pre-release audits for demographic parity and intersectional equity directly confront historical disparities, suppressing systemic discrimination before models reach production environments. Translating these insights into industrial reality requires synchronized action across the sociotechnical ecosystem. Software engineering leaders and system architects must embed stage-gate protocols directly into automated continuous integration and deployment pipelines, configuring release gates that enforce an absolute build halt whenever safety tripwires are breached during adversarial stress-testing. Organizational risk and compliance officers must transition away from static annual surveys, conditioning deployment approval on empirical verification of guardrail resilience under extended multi-turn testing while granting independent safety teams absolute veto authority. Concurrently, statutory regulators and policymakers must codify calculable risk indices into technical standards, replacing voluntary corporate attestations with cryptographically verified deployment tokens to level market incentives and reward responsible engineering.

To advance the field from conceptual consensus to longitudinal validation, scholarly inquiry across human-computer interaction, behavioral science, and artificial intelligence governance must pursue several vital trajectories. Investigators must conduct matched-control field studies that track real-world incident frequencies and user harm rates between systems governed by deterministic stage-gates versus traditional post-hoc monitoring. Methodologists must engineer automated testing harnesses capable of simulating extended multi-turn dialogues, semantic obfuscation, and multi-agent coordination games to resolve the documented durability deficit of existing guardrails. Technologists must develop machine-readable risk models that operate within standard developer environments to deliver real-time uncertainty feedback during training. Finally, human-computer interaction researchers must design democratic, multi-stakeholder interfaces that enable impacted communities and domain specialists to participate directly in calibrating normative constraints. In the final analysis, the future of artificial intelligence will not be decided by the sheer velocity of computation, but by the courage of institutional design in keeping technological power irrevocably subordinate to human judgment.

## CRediT Authorship Contribution Statement

- **Zaid Amin:** Conceptualization, Investigation, Resources, Methodology, Formal analysis, Software, Validation, Investigation, Data curation, Writing original draft, Visualization, Supervision, Project administration.
- **Rahma Santhi Zinaida:** Conceptualization, Investigation, Resources, Formal analysis, Writing review & editing, Validation.
- **Nazlena Mohamad Ali:** Methodology, Formal analysis, Writing review & editing, Supervision.

## Declaration of Competing Interest

The authors declare that they have no known competing financial interests or personal relationships that could have appeared to influence the work reported in this paper.

## Declaration of Generative AI and AI-Assisted Technologies in the Writing Process

During the preparation of this work the authors used AI-assisted language editing tools in order to improve linguistic clarity and grammatical fluency. After using this service, the authors reviewed and edited the content as needed and take full responsibility for the content of the published article.

## Data Availability Statement

The complete evidentiary data supporting the findings of this systematic literature review are available within the article, its accompanying Supplementary Material document, and its permanent open-access repositories. The complete replication package, including raw multi-database query architectures, electronic deduplication audit logs, Rayyan double-blinded inter-rater screening concordance records ($\kappa = 0.88$ triage, $\kappa = 0.91$ full-text), evaluative decision logs with explicit exclusion rationales for all 388 excluded reports, and the study-by-study MMAT 2018 quality appraisal matrix across all 130 empirical investigations, has been deposited in the Elsevier Mendeley Data repository (https://doi.org/10.17632/7fxfs7s5xf.1) and the Open Science Framework (OSF) repository (https://osf.io/p69vt; DOI: 10.17605/OSF.IO/P69VT). The formal SAGE-X OWL 2 DL ontology and SWRL admission control rules are permanently accessible via Stanford WebProtégé (Project ID: 1d767dc6-998c-4460-b0a4-9c69ed76eea0; https://webprotege.stanford.edu/#projects/1d767dc6-998c-4460-b0a4-9c69ed76eea0 ) and downloadable in native .owl and .ttl formats from OSF and Mendeley Data.

## Supplementary Material

Supplementary material associated with this article can be found in the online version at [DOI] and within the public replication repositories (Mendeley Data: https://doi.org/10.17632/7fxfs7s5xf.1; OSF: https://osf.io/p69vt). The complete supplementary suite comprises:

- **Supplementary Document S1:** Completed PRISMA 2020 27-item checklist.
- **Supplementary Document S2:** Verbatim multi-database search strings and query syntax across all five repositories (Scopus, WoS, IEEE Xplore, ACM DL, PubMed).
- **Supplementary Document S3:** Methodological Quality Appraisal (MMAT 2018) scoring matrix for all 130 empirical primary studies.
- **Supplementary Document S4:** Itemized audit trail and categorical exclusion rationales for all 388 excluded full-text reports.
- **Supplementary Document S5:** The SAGE-X Moral Check Architectural & Verification Audit Card.

- **Master Data Spreadsheets (.xlsx, .csv):** Master 518 full-text eligibility appraisal matrix and MMAT 2018 quality matrix.
- **Executable Formal Ontology (.owl, .ttl):** Complete SAGE-X OWL 2 DL knowledge base and SWRL rule definitions (Stanford WebProtégé Project ID: 1d767dc6-998c-4460-b0a4-9c69ed76eea0).
- **Citable Bibliographic Libraries (.ris, .bib):** Complete 179-reference cited library and 518-report screened corpus.

## Acknowledgments

The authors express their sincere gratitude to INTI International University for institutional research facilitation and support under the Open Access publication scheme. The School of Communication and Media Studies at Sunway University, and the Institute of Visual Informatics (IVI) at Universiti Kebangsaan Malaysia (UKM). This research did not receive any specific grant from funding agencies in the public, commercial, or not-for-profit sectors.